\documentclass[nobibnotes,nofootinbib,superscriptaddress,letterpaper]{revtex4}
\usepackage{amsmath,amssymb,bm}
\usepackage{graphicx}
\usepackage{pstricks,pst-node,pst-text,pst-3d}
\usepackage{datetime}
\usepackage{verbatim}   
\usepackage{color}      
\usepackage{subfigure}  
\usepackage{epsfig}

\def\xb{{\bf x}}

\def\rb{{\bf r}}    
\def\bb{{\bf b}}

\newcommand{\beeq}{\begin{eqnarray}}
\newcommand{\eeeq}{\end{eqnarray}}
\newcommand{\be}{\begin{equation}}
\newcommand{\ee}{\end{equation}}
\newcommand{\bea}{\begin{array}}
\newcommand{\eea}{\end{array}}

\begin{document}

\begin{abstract}
The nonlinear Balitsky-Kovchegov equation at small $x$ is solved numerically, 
incorporating  impact parameter   dependence.  Confinement is modeled by including effective gluon mass in the dipole evolution kernel, which regulates the splitting of dipoles with large sizes.  
It is shown, that the solution is sensitive to 
different implementations of the mass in the kernel.
In addition,  running coupling effects are taken into account in this analysis. 
 Finally,   a comparison of the calculations using the dipole framework with the inclusive data from HERA on the structure functions $F_2$ and $F_L$ is performed.
\end{abstract}

\title{Small x nonlinear evolution with impact parameter  and the structure function data }
\author{Jeffrey Berger}\email{jxb1024@psu.edu}
\affiliation{The Pennsylvania State University, Physics Department, University Park, PA 16802, USA}

\author{Anna M. Sta\'sto}\email{astasto@phys.psu.edu}
\affiliation{The Pennsylvania State University, Physics Department, University Park, PA 16802, USA}
\affiliation{RIKEN Center, Brookhaven National Laboratory, Upton, NY 11973, USA}
\affiliation{Institute of Nuclear Physics, Polish Academy of Sciences, ul. Radzikowskiego 152, 31-342 Krak\'ow, Poland}

\maketitle
\section{Introduction}

With particle colliders extending the  energy frontier, the need to understand QCD in the high energy limit becomes essential.  The new kinematic regime explored at the LHC and potentially in the future deep inelastic scattering (DIS) machines,  EIC \cite{Deshpande:2005wd,Boer:2011fh}  and  LHeC \cite{Dainton:2006wd,Klein:2009zz}, requires detailed analyses of the high energy and density limit in QCD.
At these high energies it is expected that the parton densities will become very large. In particular, as the Bjorken $x$ becomes very small one needs to take into account  the effects of parton  saturation \cite{Gribov:1984tu,Mueller:1985wy}.  Evolution into the small $x$ region, which incorporates parton saturation phenomenon, is governed by the nonlinear Balitsky-Kovchegov (BK) \cite{Kovchegov:1999yj,Kovchegov:1999ua,Balitsky:1995ub,Balitsky:1998ya,Balitsky:2001re,Balitsky:1998kc} equation.   A general framework which systematically incorporates the effects of the large gluon density is the 
  Color Glass Condensate (CGC) \cite{McLerran:1993ka,McLerran:1993ni} model.
  Within the CGC model the evolution into the small $x$ region is given by the 
    renormalization  group - type equation, the JIMWLK equation \cite{JalilianMarian:1997gr,JalilianMarian:1997dw,Weigert:2000gi,Iancu:2000hn,Iancu:2001ad,Ferreiro:2001qy,Mueller:2001uk}.  The CGC framework contains the BK  evolution equation as well as the evolution of the higher point gluon correlators, for a recent study  see \cite{Dumitru:2011vk}. The BK equation  is an extension 
of  the linear BFKL evolution equation for small $x$ \cite{Fadin:1975cb,Balitsky:1978ic,Lipatov:1985uk} as it takes into account parton recombination effects. These effects are included through the  additional nonlinear term in parton density. As a result the solution to this equation  cannot exceed unity, which is the  unitarity bound for the dipole (i.e. quark-antiquark pair)-target scattering amplitude.

Despite the fact that the BK equation is  closed and a relatively simple non-linear equation, no exact analytical solution yet exists. Nevertheless, there have been numerous analytical studies \cite{Levin:1999mw,Levin:2000mv,Mueller:2002zm,Munier:2003sj} as well as extensive numerical analyses \cite{Braun:2001kh,Lublinsky:2001yi,Armesto:2001fa,Lublinsky:2001bc,GolecBiernat:2001if,Rummukainen:2003ns,Albacete:2007yr}, and the properties of the solution are currently  very well known. The solution has also been used to successfully describe the experimental data on the structure function $F_2$ \cite{Lublinsky:2001yi,Albacete:2009fh,Albacete:2010sy}. Furthermore, it has been used in the prediction of a large number of processes in hadron and heavy ion collisions, such as multiplicites and single inclusive spectra \cite{Albacete:2007sm,Albacete:2010bs,Albacete:2010fs}.

In most of  these analyses one utilizes an assumption of the impact parameter independence of the solution. To be precise, the dipole target scattering amplitude in this approximation depends only on the dipole size and not on the position in impact parameter space. This leads to a significant simplification of the problem and drastically reduces the CPU time needed to numerically solve the equation.  On the other hand, the impact parameter is an important ingredient  for many of the phenomenological predictions. For example, the total multiplicities in  heavy ion collisions depend strongly on the centrality and hence knowledge of the impact parameter distribution of the partons is essential. Usually this problem is circumvented by assuming  an average parton density distribution for each of the centralities, (for a more refined approach which includes nucleon configurations in the nucleus in the initial condition see \cite{ALbacete:2010ad}). In the context of the saturation physics it is expected that the saturation scale $Q_s$,  which characterizes the dense system, is impact parameter dependent (i.e. $Q_s(b)$). The saturation scale will thus have a larger value close to the dense center of the interaction region and  smaller value at the periphery of the interaction region where the partons form a dilute system. The knowledge of the parton density distribution in impact parameter is essential not only in the context of heavy ion collisions but also for hadronic collisions (for example in the problem of the underlying event) and in particular for diffractive processes.  For example, in exclusive diffractive production of vector mesons in DIS the momentum transfer dependence crucially depends on the  impact parameter profile of the dipole scattering amplitude \cite{Munier:2001nr,Kowalski:2003hm,Kowalski:2006hc} 

The impact parameter dependence has been taken into account in Monte Carlo simulations based on dipole evolution and scattering \cite{Salam:1995zd,Salam:1995uy,Avsar:2005iz,Avsar:2006gw,Avsar:2006jy,Avsar:2007xh} as well as in numerical solutions to the BK evolution equation, 
    see \cite{GolecBiernat:2003ym,Berger:2010sh} and \cite{Gotsman:2004ra}.   It has been also discussed in the context of the conformal properties of the equation \cite{Gubser:2011qva}.
 The results so far indicate that there are important modifications to the solution when the impact parameter is taken into consideration. In particular, it has been found that the parton (or more precisely dipole) density distribution in impact parameter  possesses long range Coulomb-like power tails \cite{Salam:1995uy,Salam:1995zd,GolecBiernat:2003ym,Berger:2010sh}, which are the direct consequence of the form of the perturbative branching kernel.  These tails need to be regularized by the appropriate cuts on the large dipole sizes which mimic confinement effects \cite{Gotsman:2004ra,Avsar:2005iz,Avsar:2006jy,Avsar:2006gw,Avsar:2007xh}. The other important observation was that the dipole scattering amplitude  decreases with increasing dipole size at large dipole sizes (when it is evaluated at fixed value of impact parameter). This has to be contrasted with the impact parameter independent solutions for which the amplitude always saturates to unity for arbitrarily large  values of the dipole size.

In the earlier work \cite{Berger:2010sh} we explored the  dynamics of the BK equation with impact parameter by taking into account subleading effects such as kinematical cuts and the running of the strong coupling. 
The goal of this paper is to use the solutions to the BK equation with impact parameter dependence to compute the cross sections and structure functions for the deep inelastic process. 
In order to perform this analysis we introduce the initial condition with physical scales and also modify the branching kernel to account for non-perturbative confinement effects.
This is done by introducing a mass parameter $m$ into the kernel which modifies the long distance behavior of the dipole-target scattering amplitude.  This parameter restricts the splitting of dipoles into daughter dipoles which are larger than $r_{\rm max}=\frac{1}{m}$ in order to account for confinement. With such a setup we then compute
$F_2$ and $F_L$ structure functions using the resulting solutions and compare them with experimental data from HERA \cite{Collaboration:2010ry,:2009wt}. 

The resulting dipole scattering amplitude was then compared with the parametrizations
available in the literature \cite{Kowalski:2006hc} which include impact parameter.
 In particular, we find that although the  
dynamically generated amplitude from the BK equation is similar for small values of the dipole size to the Glauber-Mueller like parametrization, for larger dipole sizes one observes notable differences. To be precise, we find that the BK equation generates  solutions which possess specific correlations between the dipole size and impact parameter, an effect which is totally absent in the Glauber-Mueller type parametrizations.

The outline of the paper is the following: in Sec.~\ref{sec:CSdipole} we state the basic formulae for the inclusive cross section within the framework of the dipole model.  In Sec.~\ref{sec:Evol} we briefly discuss the most substantial features of the solution with  the impact parameter dependence and discuss the differences with respect to the impact parameter independent scenario. We then introduce modifications to the evolution kernel  by including the mass which mimics confinement. We discuss the properties of the solutions which result from these modifications.  Both fixed coupling and running coupling cases are considered as well as the various methods that we used to implement the mass parameter which regulates the large dipole sizes.  Later, in Sec.~\ref{sec:results} we show the results for  the dipole amplitude and we make the first comparison  to the data for $F_2$ and $F_L$.  Finally,  in Sec.\ref{sec:conclusion} we state the conclusions.

\section{DIS inclusive cross section within the dipole model}
\label{sec:CSdipole}

\begin{figure}
\centering
\subfigure[]{\label{fig:dipolemodel1}\includegraphics[angle=0,width=0.4\textwidth]{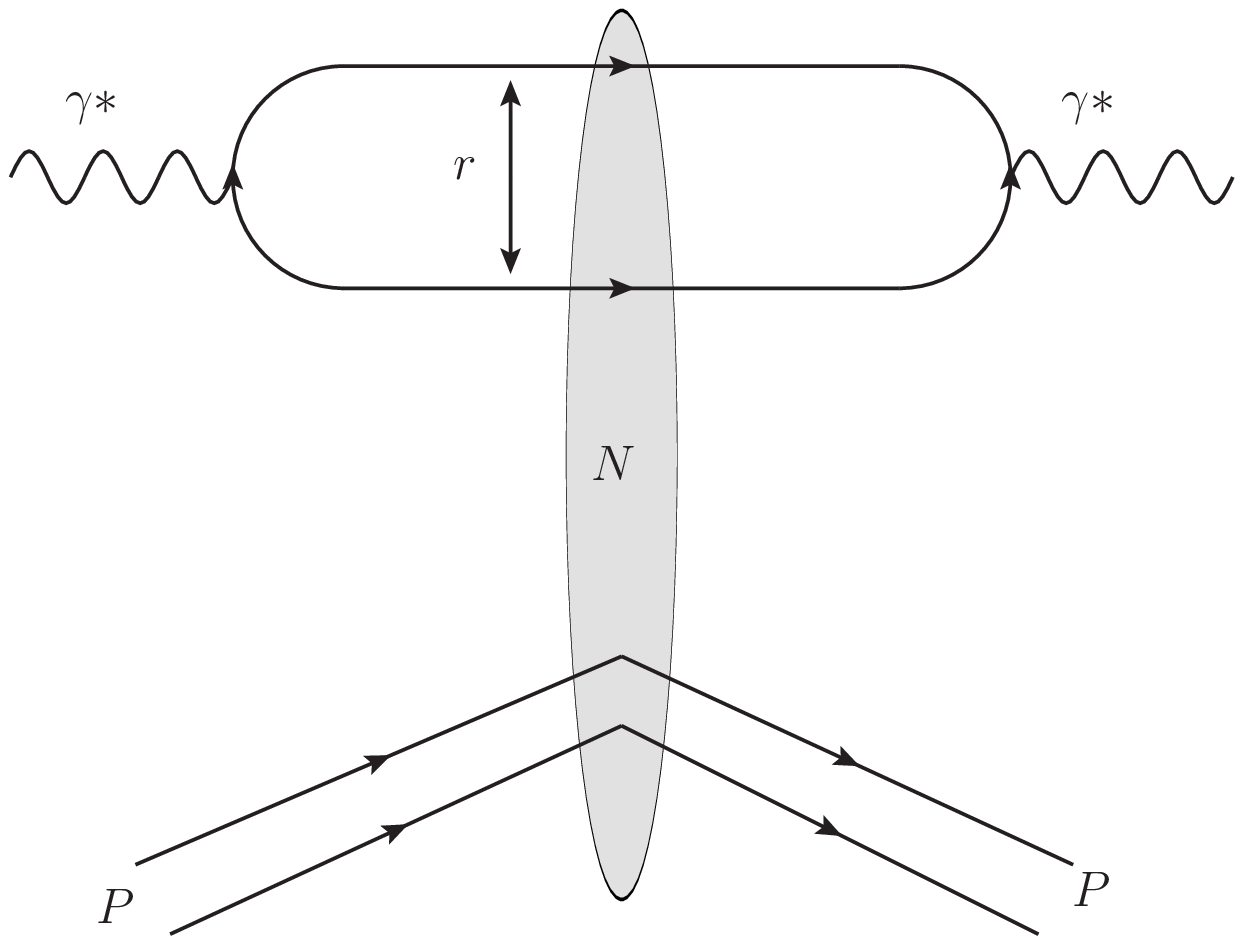}}
\subfigure[]{\label{fig:dipolemodel2}\includegraphics[angle=0,width=0.4\textwidth]{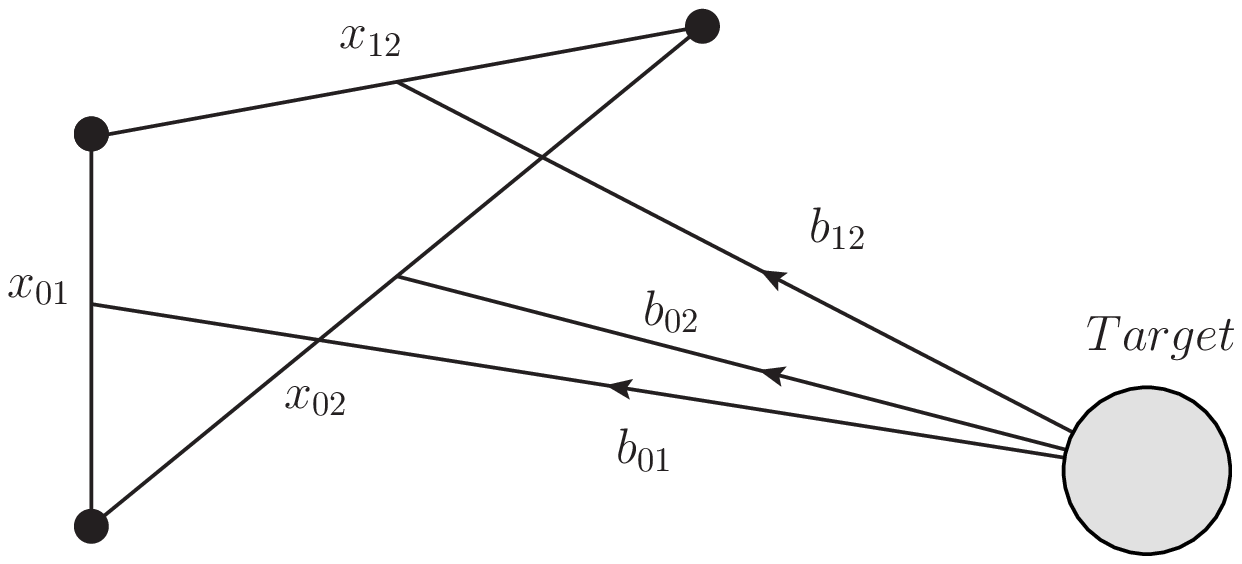}}
\caption{ Plot (a): schematic representation of the dipole  picture in DIS.  The incoming virtual photon $\gamma^*$ splits into a color dipole (quark-antiquark pair) of size $r$ which subsequently  interacts with a target, where $N$ is the dipole-target  scattering amplitude.  Plot (b):  depiction in transverse space of single branching of the parent dipole ${01}$ into two daughter dipoles. The sizes of dipoles are denoted by $x_{ij}$. The  impact parameter variables $b_{ij}$ of all the dipoles are shown relative to the target.}
\label{fig:dipolemodel}
\end{figure}

The dipole model \cite{Nikolaev:1990ja,Nikolaev:1991et} is a very useful tool in evaluating many processes at small values of $x$.  One of the advantages of this approach is the possibility of including multiple parton scattering effects. It has been originally formulated for the description of deep inelastic lepton-proton (or nucleus) scattering at small $x$. In this picture, utilizing the leading logarithmic approximation in $x$, the incoming electron emits a virtual photon which fluctuates into a quark-antiquark pair, a dipole.  The color dipole then subsequently interacts with the parton constituents of the nucleon, as  is illustrated in Fig.~\ref{fig:dipolemodel1}. The interaction of the dipole pair with the target is given by the scattering amplitude $N$.   The $q\bar{q}$ pair is characterized by a dipole size which is defined as a separation distance of the color charges $\xb_{01} = \xb_0 - \xb_1$ (where $\xb_0$ and $\xb_1$ are the positions of the $q$ and $\bar{q}$ in transverse space).\footnote{  In this paper we shall denote vector quantities in bold, otherwise they should be read as magnitudes of the associated vector.  Also, alternatively we will be also using here the notation for the dipole size  to be  $r=x_{01}$ and impact parameter $b=\frac{|\xb_0 + \xb_1|}{2}$. } The transverse momentum of the quarks in the dipole  is of the order of $\sim \frac{1}{x_{01}}$ where large dipoles correspond to the infra-red region and need to be regulated, as  will be discussed in detail in later sections.  The interaction of the dipole with the target is described by the scattering amplitude $N(\rb,\bb;Y)$ which contains all the information about the dynamics of the strong interaction. In the following analysis the   full dependence of the scattering amplitude on the impact parameter $\bb$  will be taken into account.  The evolution of the amplitude in rapidity $Y$ can be represented as emissions of daughter dipoles from the original parent dipole.  When  the original dipole ${01}$ splits into two dipoles $02$ and $12$ a new coordinate appears, $\xb_2$.  These two daughter dipoles are produced  with sizes $\xb_{12}$ and $\xb_{02}$ at impact parameters $\bb_{12}$ and $\bb_{02}$ as illustrated in Fig.~\ref{fig:dipolemodel2}.

The dipole-target amplitude $N(\rb,\bb;Y)$ at high values of rapidity $Y$ (or small $x$) is found from the solution to the BK evolution equation which can be represented in the following form:
\be
\frac{\partial N_{\xb_0\xb_1}}{\partial Y} = \int\frac{d^2\xb_2}{2\pi}\, {\cal  K}(x_{01},x_{12},x_{02}; \alpha_s,m)\left[N_{\xb_0\xb_2}+N_{\xb_2\xb_1}-N_{\xb_0\xb_1}-N_{\xb_0\xb_2}N_{\xb_2\xb_1}\right] \; .
\label{eq:BK}
\ee
In the above equation we used the shorthand notation for the arguments of the amplitude $N_{\xb_i\xb_j}\equiv N({\rb_{ij}=\xb_i-\xb_j},\bb_{ij}=\frac{1}{2}(\xb_i+\xb_j);Y)$
which  depends on the two transverse positions $\xb_i$ and $\xb_j$ and on the rapidity $Y$.  The branching kernel ${\cal K}(x_{01},x_{12},x_{02}; \alpha_s,m)$ depends on the dipole sizes involved and contains all information about the splitting of the dipoles. In addition it depends on  the running coupling $\alpha_s$. We have also indicated that it depends  on the infra-red cutoff $m$ imposed on large  dipoles.

 The solution to Eq.~(\ref{eq:BK}) is the dipole-target scattering amplitude for arbitrarily small $x$.  In order to compute the 
 structure functions $F_2$ and $F_L$ for the proton we use the following standard formulae in the dipole picture in the transverse coordinate representation

\be
F_2(Q^2,x) = \frac{Q^2}{4 \pi^2 \alpha_{em}}\int{d^2 {\bf r} \int_0^1 dz \left(|\Psi_T(r,z,Q^2)|^2+|\Psi_L(r,z,Q^2)|^2\right) \sigma_{\rm dip}({\bf r},x)} \; ,
\label{eq:F2}
\ee
and
\be
F_L(Q^2,x) = \frac{Q^2}{4 \pi^2 \alpha_{em}}\int{d^2 {\bf r} \int_0^1 dz |\Psi_L(r,z,Q^2)|^2 \sigma_{\rm dip} (\rb,x)} \; .
\label{eq:FL}
\ee
Here, $\sigma_{dip}$ is the (dimensionful) dipole cross section obtained from the (dimensionless) scattering amplitude by integrating over the impact parameter
\be
\sigma_{\rm dip}(\rb,x) = 2 \int{d^2\bb \, N(\rb,\bb;Y)} \; ,  \; \; \; \; \; \;Y=\ln 1/x \; .
\label{eq:sigmadip}
\ee

The $\Psi(\rb,Q^2,Y)_{T/L}$ functions are the  photon wave functions. They describe the   dissociation of a photon into a $q$$\bar{q}$ pair and can be calculated from perturbation theory. The photon wave function has the following form for the case of transverse photon polarization
\be
|\Psi_T(r,z,Q^2)|^2 = \frac{3 \alpha_{em}}{2 \pi^2} \sum_f e_f^2\left(\left[z^2 + (1-z)^2\right]\bar{Q}^2_f K_1^2\left(\bar{Q}_f r\right) + m_f^2 K_0^2\left(\bar{Q}_f r\right)\right) \; ,
\label{eq:PhotonT}
\ee
and for longitudinal polarization
\be
|\Psi_L(r,z,Q^2)|^2 = \frac{3 \alpha_{em}}{2 \pi^2} \sum_f e_f^2\left(4Q^2z^2(1-z)^2K_0^2\left(\bar{Q}_f r\right)\right) \; .
\label{eq:PhotonL}
\ee
In the above equations  $\bar{Q}^2_f = z(1-z)Q^2 + m_f^2$, where $-Q^2$ is the photon virtuality and $z,(1-z)$ are the fractions of the longitudinal momentum of the photon carried by the quarks. In addition  $K_{0,1}$ are modified Bessel functions of the second kind.  The summations are over the active quark flavors $f$ of charge $e_f$ and mass $m_f$.

\section{Dipole evolution with impact parameter dependence}
\label{sec:Evol}

The solution to the  BK equation with impact parameter dependence is found  numerically, the description of the procedure was outlined in \cite{GolecBiernat:2003ym,Berger:2010sh}. 
The technical complication when trying to solve BK with impact parameter is the increased number of arguments in the dipole amplitude. In the $b$-independent scenario the amplitude depends only on two variables: rapidity and dipole size. 
In the $b$-dependent case there
are  5 variables: rapidity, dipole size (vector, 2-dim.) and impact parameter (vector, also 2-dim.). Alternatively, one can choose the coordinate variables to be parametrized by the dipole size, impact parameter and two angles: one parametrizing the absolute orientation of the dipole-target system in the coordinate space and the second  describing the relative orientation of the dipole with respect to the target. 
With the assumption of the global rotational symmetry (i.e. independence of the first angle) the number of variables reduces to 4: rapidity, dipole size, impact parameter and the angle between $\rb$ and $\bb$. Such large number of variables requires working with a very large multidimensional grid and it leads to a significant increase of computation time per each step of evolution in rapidity.  
The BK equation was solved numerically by discretizing the scattering amplitude in terms of  variables  $(\log_{10}r,\log_{10}b,\cos \theta)$, where $\theta$ is the angle between impact parameter $\bb$ and dipole size $\rb$.    The amplitude $N(r,b,\cos\theta)$ was placed on a grid with dimensions $200_r\times200_b\times20_\theta$. More details can be found in Refs.~\cite{GolecBiernat:2003ym,Berger:2010sh}.

\begin{figure}
\centering
\subfigure[ \hspace*{0.1cm} Small impact parameter: $b=1$. ]{\label{fig:OldData1}\includegraphics[angle=270,width=0.49\textwidth]{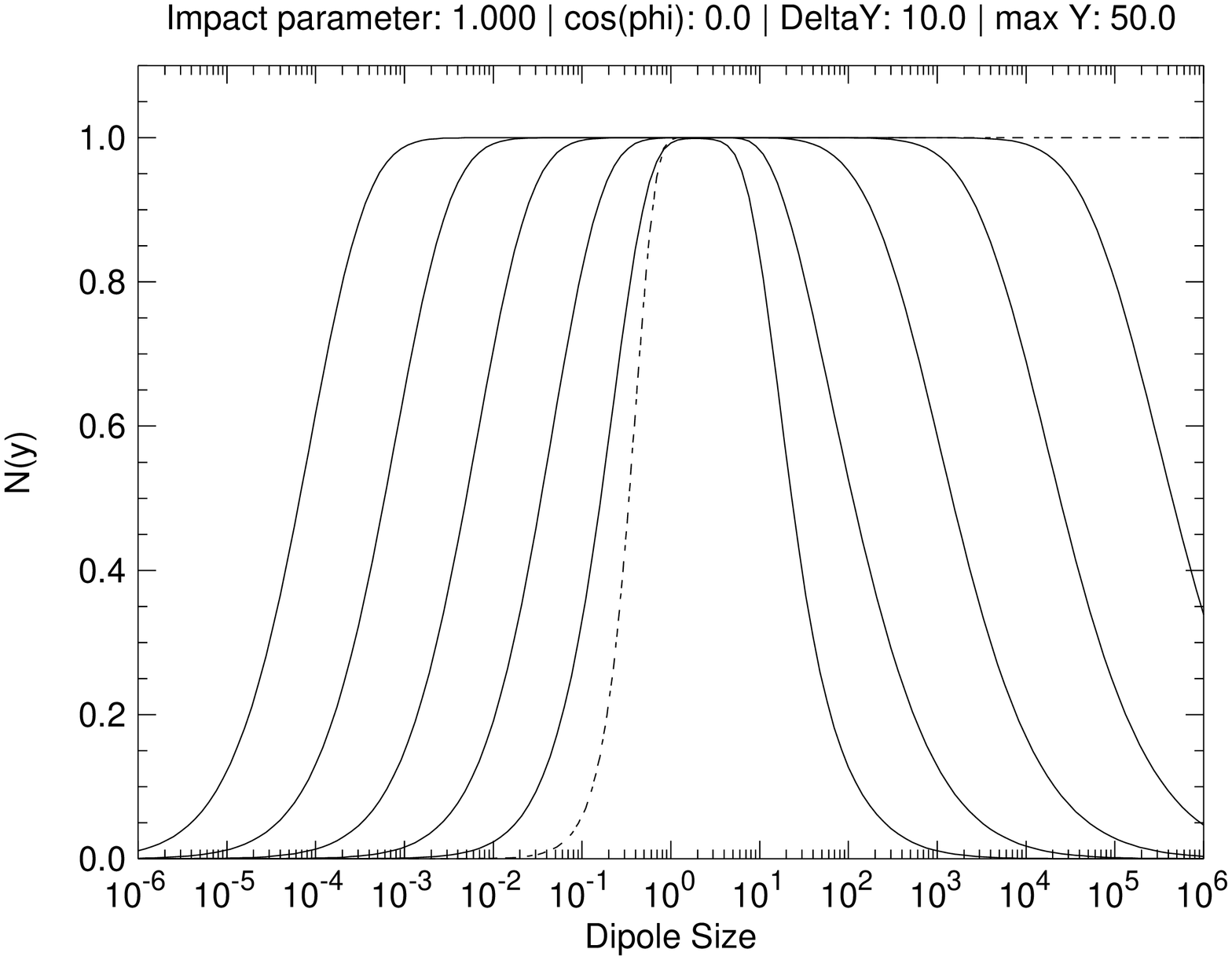}}
\subfigure[ \hspace*{0.1cm} Large impact parameter: $b=100$.]{\label{fig:OldData2}\includegraphics[angle=270,width=0.49\textwidth]{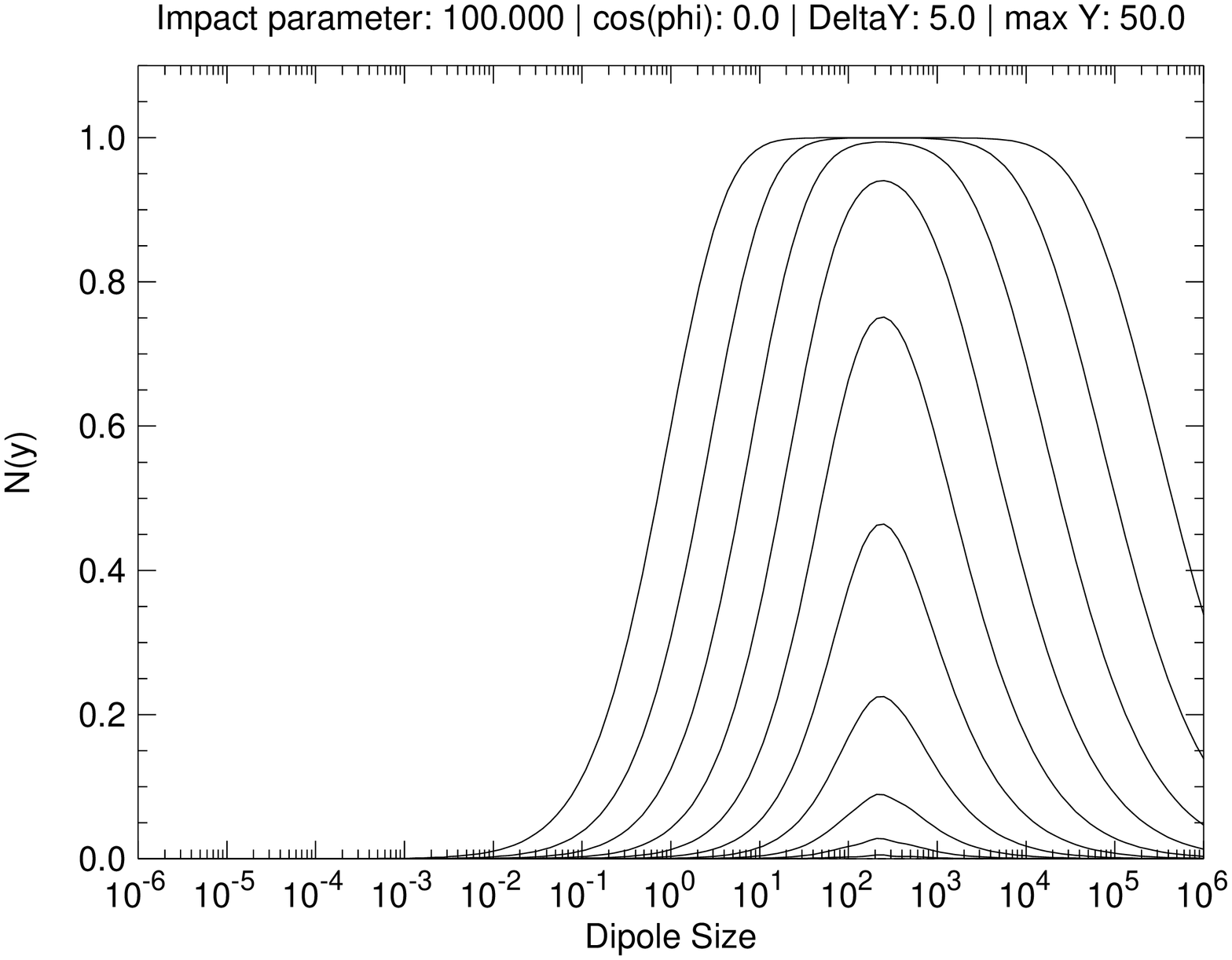}}
\caption{The dipole scattering amplitude $N(\rb,\bb;Y)$ as a function of the dipole size $r$  from the solution to the LL BK equation with impact parameter dependence. The strong coupling is fixed $\bar{\alpha}_s=0.2$. The consecutive curves shown in plots are for rapidities $Y=10,20,30,40,50$ on the  plot (a) and in  rapidity intervals of $5$, until $Y=50$ on  plot (b). The dotted-dashed line in the  left plot is the initial condition at $Y=0$ given by Eq.~\ref{eq:n0simp}. The initial condition is not visible on the right plot as it is very close to zero.  The orientation of the dipole with respect to the target is such that ${\rb} \perp {\bb}$.}
\label{fig:OldData}
\end{figure}

Let us  briefly summarize the most important  results of \cite{GolecBiernat:2003ym,Berger:2010sh} which pertain to the properties of the solution with the impact parameter and the differences with respect to the impact parameter independent approximation. We note that for the solutions presented in Fig.~\ref{fig:OldData} we used the same initial condition at $Y=0$ as in Ref.~\cite{Berger:2010sh}, which was taken to be
\begin{equation}
N^0=1-\exp(-c_1 r^2 \exp(-c_2 b^2)) \; ,
\label{eq:n0simp}
\end{equation}
with $c_1=10, \, c_2=0.5$.

\begin{figure}
\centering
\subfigure[ \hspace*{0.1cm}Small dipole scattering.]{\label{fig:DipoleTarget1}\includegraphics[width=0.17\textwidth]{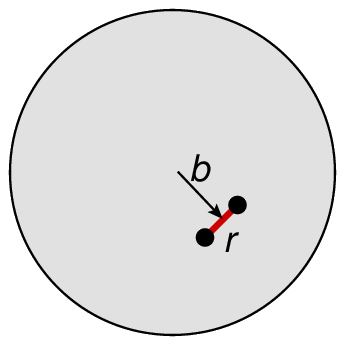}}
\hspace*{4cm}
\subfigure[ \hspace*{0.1cm}Large dipole scattering.]{\label{fig:DipoleTarget2}\includegraphics[width=0.24\textwidth]{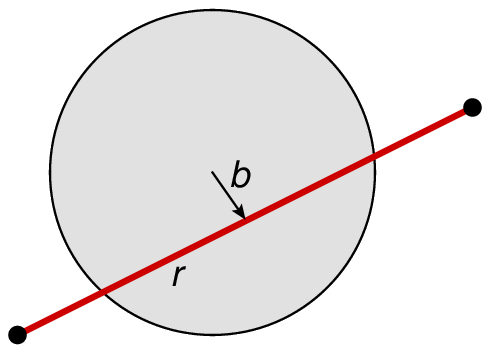}}
\caption{Two different dipole-target configurations for which the dipole scattering amplitude is small (away from unitarity limit). }
\label{fig:DipoleTarget}
\end{figure}

The most distinctive feature of the solution with impact parameter dependence is that  the amplitude for large dipole sizes   goes to zero. This  is clearly illustrated in Fig.~\ref{fig:OldData1} where the solution for fixed value of impact parameter $b$ is shown as a function of the dipole size $r$. This property of the amplitude has to be contrasted with the impact parameter independent solution, in which case the amplitude is always equal to unity for sufficiently large dipole sizes.  The fact that the $b$-dependent amplitude drops for large dipole sizes has a rather simple physical interpretation: the interaction region possesses finite extension in impact parameter space. The size of this  region is set at low rapidity by the initial conditions (in our case it is parameter $c_2$ in Eq.~\ref{eq:n0simp}) and is later increased in the course of the evolution by the diffusion of the dipoles in transverse coordinate space. The probability of the scattering for  dipoles which have sizes larger than the extension of the  interaction region is very small  and therefore the amplitude will tend to go to zero for such configurations.
 The dipole-target amplitude therefore is  largest for the scattering of dipoles with sizes  comparable with  the typical size of the target. 
 As a result, the amplitude decreases either for small or for very large dipole sizes, which in each case are very different than the extension of the target, yielding a maximum contribution for some intermediate sizes of dipoles. The configurations for which the amplitude is small are schematically illustrated in Fig.~\ref{fig:DipoleTarget}, and they correspond to the tails of the distribution depicted in Fig.~\ref{fig:OldData1}.  In the course of the evolution in rapidity this distribution in the dipole size broadens due to diffusion. As a consequence, the amplitude in the impact parameter dependent scenario has two fronts as is evident from Fig.~\ref{fig:OldData1}.  The first front (for small dipoles) is similar to the front in the impact parameter independent case, and one can define the saturation scale which divides the dense (i.e. where $N\sim 1$ ) and dilute ( $N \ll 1$ ) regime for small dipoles. This saturation scale can be parametrized in  the form  $Q_{sL}^2(Y,b)=Q_{0,sL}^{2} \exp( \lambda_{sL} Y)$ with $\lambda_{sL}\simeq 4.4$ which is consistent with the analytical predictions and with the solutions for the impact parameter independent case. 
 The second front expands towards larger dipoles with increasing rapidity.
 One can also define the corresponding 'saturation scale' for large dipoles which can be similarly parametrized as $Q_{sR}^2(Y,b)= Q_{0,sR}^{2}\exp(- \lambda_{sR} Y)$ with  $\lambda_{sR}\simeq5.8$. Note the '-' sign in the exponent, which originates from the fact that the second front expands towards larger dipoles with increasing rapidity, and therefore this second 'saturation scale'  decreases with rapidity. One has to stress however that the region of large dipoles is going to be heavily modified by the non-perturbative effects (see the discussion later in this section) and therefore this second saturation scale is most likely only an artefact of the perturbative expansion in the leading logarithmic (LL) approximation.

 This novel feature of the solution in impact parameter dependent case, namely the decrease of the amplitude at large dipole sizes is directly related to the profile in the impact parameter space. It was found in the LL case \cite{GolecBiernat:2003ym,Berger:2010sh} that the solution  has a power-like tail in impact parameter. This is related to the fact that there are no mass scales in the perturbative evolution and hence the interaction is long range \cite{Kovner:2001bh,Kovner:2002yt}. As we will discuss later in this section, the branching kernel in the equation needs to be modified by including the effective gluon mass in order to regulate the power-like behavior of the amplitude for large dipole sizes and include the effects of confinement.

Another distinctive feature of the solutions with impact parameter dependence is the presence of the strong correlations between the dipole size and the impact parameter.
It has been observed that the amplitude is largest for specific configurations and orientations of the dipole size  ${\rb}$ and impact parameter ${\bb}$ vectors. In particular, the amplitude has a peak 
 when the dipole size is equal  twice the impact parameter. 
 This is clearly illustrated in 
 Fig.~\ref{fig:OldData2}  where the peak in the amplitude at $r=2b$ emerges.
 In this case there is also a non-negligible dependence on the angle between the dipole size vector and the impact parameter.
 It turns out that the amplitude is largest when the angle  between the dipole size vector and the impact parameter vector is equal to $0$ or $\pi$, that is when the vectors ${\rb}$ and ${\bb}$ are parallel or anti-parallel. In this configuration  one of the color charges scatters off the center of the target.  This enhancement can also be seen  analytically from the conformal eigenfunction representation as discussed in \cite{Berger:2010sh}.

By integrating the resulting amplitude $N(\rb,\bb;Y)$ over the impact parameter $\bb$, as in Eq.~\ref{eq:sigmadip}, one obtains the dipole cross section as a function of the dipole size and rapidity. Even though the amplitude $N$ never exceeds unity, the resulting dipole cross section increases very strongly with the rapidity \cite{Kovner:2001bh,Kovner:2002yt}, due to the rapid diffusion of dipoles in the impact parameter space. In the leading logarithmic approximation this increase is exponential in rapidity, $\sigma_{\rm dip}\sim \exp(\lambda_B Y) $, with $\lambda_B\simeq 2.6$ ~\cite{GolecBiernat:2003ym,Berger:2010sh}. This behavior is very different from the features observed in the $b$-independent solution.  
In the latter case one assumes that the dipole amplitude does not depend on the position in coordinate space, but only on the absolute value of the dipole size and rapidity, $N(r;Y)$. The solution still saturates to unity which is the fixed point of the equation in this approximation as well. To obtain the dipole cross section one needs then to multiply  the amplitude by a dimensionful coefficient i.e.
$$
\sigma_{\rm dip} \; = \; {\cal S}_0 \; N(r;Y) \; ,
$$
where ${\cal S}_0$ can be interpreted as the integral 
$$
{\cal S}_0=\int_{\cal R} d^2 \bb
$$
over  the interaction region ${\cal R}$ in the impact parameter and is entirely introduced by hand.  The behavior of ${\cal S}_0$ on rapidity (whether it is constant or increasing) is thus not determined by the $b$-independent BK evolution equation. 

We see therefore, that the $b$-dependent solution to the BK equation, when integrated over the impact parameter does not reduce to the solution in the previously studied approximation when the impact parameter is neglected. This is an important qualitative and quantitative difference between these two solutions. In fact, the solution with impact parameter dependence is more physically motivated as it gives the increase of the dipole cross section with rapidity. It will also naturally lead to the  increase  with the energy of the diffractive slope for the vector meson production. We note however, that the results in the LL approximation, even with the running coupling, are not compatible with the experimental data. This is due to the fact  that the diffusion in impact parameter is very strong (since it is driven by the purely perturbative physics) and results in a fast increase of the interaction radius which is not supported by the data. In order to tame this growth further corrections are necessary, in particular the inclusion of subleading corrections and  a mass parameter into the evolution as will be discussed later in this section.

\subsection{Initial condition for the evolution towards small $x$}
\label{sec:initial}
For the rest of the numerical simulations presented in this paper we will use the initial conditions of the similar form as in (\ref{eq:n0simp}) but with parameters which were adjusted to obtain the predictions consistent with experimental data. We will use as our initial condition the parametrization from Ref.~\cite{Kowalski:2003hm}, where the parametrization in the Glauber-Mueller form was used

\be
N_{\rm GM}(r,b;Y=\ln 1/x) \, = \, 1 - \exp{\left(-\frac{\pi^2}{2N_c}r ^2 x g(x,\eta^2) T(b)\right)} \; ,
\label{eq:glaubermueller}
\ee
with 
\be
T(b) \, = \, \frac{1}{8 \pi} e^{\frac{-b^2}{2B_G}} \; .
\label{eq:profile} 
\ee
The formula (\ref{eq:glaubermueller}) is used as an initial condition for the BK evolution for dipoles with sizes smaller than the cutoff $1/m$. For dipoles larger than the cutoff the initial condition is set to zero ( see the discussion in the next subsection).
In formula (\ref{eq:glaubermueller}) the function $xg(x,\eta^2)$ is the integrated gluon density function and $T(b)$ is the density profile of the target in transverse space with $B_G=4 \; {\rm GeV^{-2}}$. This parameter was  set to fit the $t$ slope of the diffractive $J/\Psi$ production in \cite{Kowalski:2006hc}.   Also, the  scale $\eta$ was set according to \cite{Kowalski:2006hc} to be equal to $\eta^2=\frac{C}{r^2}+\eta_0^2$ with parameters $C=4$ and $\eta_0^2=1.16 \; {\rm GeV}^2$. The  integrated gluon density in (\ref{eq:glaubermueller}) was also taken from fits performed \cite{Kowalski:2006hc}. We use (\ref{eq:glaubermueller}) as the initial condition at $Y_0=\ln 1/x_0$, $x_0=10^{-2}$ and evolve the amplitude with the BK equation to obtain the solution at lower values of $x<x_0$. 
We also note that the initial condition (\ref{eq:glaubermueller}) depends only on the absolute values of the dipole size and impact parameter. The nontrivial dependence on the angle between vectors $\rb$ and $\bb$ is not present in the initial condition, instead being dynamically generated when the initial condition is evolved with the BK equation.


\subsection{Including the effective gluon mass into the evolution kernel}

Currently it is unknown  how to introduce a massive cutoff on a fundamental level into the small $x$ evolution as it is an entirely non-perturbative problem. We have tested various prescriptions  and found that there is a rather large sensitivity of the resulting solutions to the details of confinement implementation. In addition, there is a strong dependence of the solutions  on the way the running coupling is regularized. 
This sensitivity stems from the  behavior of the $b$-dependent solution at large dipole sizes, as discussed previously. One important difference to note between the solution with and without the impact parameter dependence is that in the latter case the running of the  strong coupling is naturally regularized by the saturation scale, provided the latter is in the semi-hard regime. For example, it was observed  in Ref.~\cite{GolecBiernat:2001if}  that different prescriptions of the regularizations for the running coupling gave similar results in the case of the non-linear evolution. The  emergence of the semi-hard saturation scale $Q_s$ and its role as an infrared cutoff  is one of the most prominent and useful features of the nonlinear evolution. In the case when the impact parameter is taken into account, the saturation scale  strongly varies with $b$. In the dense region this scale is large, and is providing a natural cutoff for the running coupling in the same way as in the impact parameter independent solutions.  There is however always a peripheral interaction region in impact parameter where the scattering amplitude is small and  the system is dilute.  Consequently the coupling is not regularized in this region by the saturation scale which is very small for large $b$.  As a result,   the solution in the dilute peripheral regime is  governed by the linear evolution and  becomes extremely sensitive to the region of  large dipole sizes.

The specific value of the massive cutoff which is implemented into the evolution kernel   should correspond to the non-perturbative scale which is  related to confinement. Lattice simulations \cite{Dudal:2010tf,Oliveira:2010xc} suggest  the presence of an effective gluon mass which would regulate the large-dipole size regime.  In the case of BK equation an important ingredient  that one has  to take into account is the fact that the evolution in impact parameter is strongly correlated with the evolution of the dipole sizes. 
Therefore   the cutoff on the latter will crucially influence the size of the interaction region in impact parameter and its variation with the collision  energy. 
In the Monte Carlo analysis of dipole evolution in \cite{Flensburg:2008ag,Avsar:2006jy} the parameter $r_{\max}$ was set to be around $3 \; {\rm GeV^{-1}}$ (to be precise it was set to  $2.9 \; {\rm GeV^{-1}}$ in \cite{Flensburg:2008ag} and $3.1 \; {\rm GeV^{-1}}$ in \cite{Avsar:2006jy}). We set the value of the cutoff to be of the same order, i.e.$m=\frac{1}{r_{\rm max}}=0.35 \; {\rm GeV}$ which corresponds to $r_{\rm max}\simeq 2.86 \; {\rm GeV^{-1}}$.

Let us finally note here
that the  impact parameter profile  can be accessed through the measurement of the diffractive production of the vector mesons \cite{Kowalski:2006hc,Kowalski:2003hm}. 
From the experimental data \cite{Aktas:2005xu}  at $|t|<1.2 \; {\rm GeV}^2$ it is known that the diffractive slope  $B_D$ of the $J/\Psi$ production ($\frac{d\sigma}{dt}\sim e^{B_D t}$)  is of the order of $\sim 4.57 \; {\rm GeV}^{-2}$ for $Q^2 \lesssim 1 \; {\rm GeV}^2$ and  $\sim 3.5 \; {\rm GeV}^{-2}$ for $Q^2>5 \; {\rm GeV}^2$ in the energy range $40 < W_{\gamma p}<160 \; {\rm GeV}$. It is also slowly growing with the increasing energy $W$ of the $\gamma^* p$ system.
The value of $r_{\rm max}$ we set in the calculation will strongly influence the variation of the width of the impact parameter profile with the energy. Therefore in principle $r_{\rm max}$ can be related to   the dependence of the $B_D$ with the energy, \cite{BerStaDiff}.

\subsubsection{ Regularization of the kernel  at leading logarithmic accuracy with fixed coupling}
\label{sec:LOKernel}

The regularization of the large dipole sizes in the dipole kernel is a  purely non-perturbative effect. One possible implementation, which is physically motivated,  is to introduce the effective gluon mass $m$   into the propagators. This mass will set a correlation length $r_{\rm max}\equiv1/m$ which will limit the propagation of the strong color force. One can then derive the dipole  kernel by computing the emission of the gluon from the initial $q\bar{q}$ dipole \cite{Nikolaev:1993th,Nikolaev:1993ke}.   The Fourier transform of the momentum space expression of the $q\bar{q}g$ lightcone density into the coordinate space results in the expression which contains  the modified Bessel functions instead of powers as in the LL expression.  This means that instead of the long range Coulomb-type interaction present in the perturbative evolution there is  now  screened Yukawa force with finite range given by $r_{\rm max}$. The modified branching kernel for dipoles with effective gluon mass $m$ has the following form \cite{Nikolaev:1993th,Nikolaev:1993ke}
\be
{\cal K}_{LL}^{(1)} \, =\,  \frac{{\alpha}_s N_c}{\pi} \,m^2 \left[ K_1^2(m x_{02})+K_1^2(m x_{12}) - 2 K_1(m x_{02}) K_1(m x_{12})\frac{\xb_{02} \cdot \xb_{12}}{x_{02} \, x_{12}}\right] \;.
\label{eq:KernLOBessMass}
\ee
In the limit when the dipole sizes are small compared to the cutoff,  $x_{ij} \ll \frac{1}{m}$, the Bessel functions are approximated as  $K_1(m x_{ij})\simeq \frac{1}{mx_{ij}}$. In this limit  the kernel  (\ref{eq:KernLOBessMass}) obviously reduces to the well known expression in the LL approximation \cite{Mueller:1993rr,Nikolaev:1993th,Nikolaev:1993ke}. On the other hand, the production of large dipoles $x_{ij} \gtrsim \frac{1}{m}$ is exponentially suppressed.
This form of the modification of the kernel (\ref{eq:KernLOBessMass}) was also used in the later version of the  Monte Carlo simulation  for the dipole splitting  and evolution \cite{Flensburg:2008ag}.

By inspecting expression (\ref{eq:KernLOBessMass}) it is clear that this kernel does not vanish  completely when one of the dipole sizes  is larger than $1/m$ but the other is smaller than $1/m$.  In other words even for very large parent dipole size the above kernel permits the splitting, provided one of the daughter dipoles  sizes is below the cutoff. It means that  despite the presence of the cutoff $1/m$ there is still  a diffusion into the region of arbitrarily large dipole sizes. 

As an alternative to the above scenario we have thus used a second prescription where the splitting is suppressed whenever any of the daughter dipole sizes is larger than the cutoff. To be precise, we have tried the second ansatz of the form

\be
{\cal K}_{LL}^{(2)} \,=\,   \frac{{\alpha}_s N_c}{\pi} \frac{x_{01}^2}{x_{02}^2x_{12}^2} \; \Theta(\frac{1}{m^2} - x_{02}^2)\Theta(\frac{1}{m^2} - x_{12}^2) \; .
\label{eq:KernLOTheta}
\ee

The kernel is thus set to zero whenever any of the  dipoles is larger than $1/m$.
This form gives more suppression of the dipole splitting than the first modification (\ref{eq:KernLOBessMass}) and results in an overall slower evolution in rapidity.

\subsubsection{Regularization in the presence of the running coupling }
\label{sec:rc}

The  implementations of the cutoff presented above can be used  with a fixed coupling. The running coupling correction brings in an additional complication as the form of the kernel changes rather significantly and the coupling is no longer a simple multiplicative factor.  The running coupling corrections to the BK evolution have been computed in two independent calculations \cite{Balitsky:2006wa} and \cite{Kovchegov:2006vj}.  The two schemes differ by the non-trivial subtraction term as was demonstrated in detail in \cite{Albacete:2007yr} that we will not consider in this work.  
For the purpose of the current work we use the scheme of \cite{Balitsky:2006wa} 
\be
{\cal K}_{rcLL}^{\rm Bal} = \frac{\alpha_s(x_{01}^2) N_c}{\pi} \left[\frac{1}{x_{02}^2}\left(\frac{\alpha_s(x_{02}^2)}{\alpha_s(x_{12}^2)} - 1\right) + \frac{1}{x_{12}^2}\left(\frac{\alpha_s(x_{12}^2)}{\alpha_s(x_{02}^2)} - 1\right) + \frac{x_{01}^2}{x^2_{12} x_{02}^2}\right]
\label{eq:KernLOBal} \;.
\ee
The other scheme derived \cite{Kovchegov:2006vj} tends to be  more time consuming per one evaluation which in the case of the impact parameter dependent simulations with large grids greatly extends the time necessary for the evolution. Therefore for purely practical reasons we utilized only Balitsky prescription (\ref{eq:KernLOBal}). 

Kernel (\ref{eq:KernLOBal}) reduces to the  LL kernel with the strong coupling evaluated at the smallest value of the dipole size, for configurations of dipole sizes which are strongly ordered. For example, in the case when $x_{01} \sim x_{02} \gg x_{12}$ we have
\begin{equation}
{\cal K}_{rcLL}^{\rm Bal} \simeq 
 \frac{ \alpha_s(x_{01}^2) N_c}{\pi} \left[ \frac{1}{x_{12}^2}\left(\frac{\alpha_s(x_{12}^2)}{\alpha_s(x_{02}^2)} - 1\right) + \frac{1}{x^2_{12} }\right]\simeq \frac{N_c \alpha_s(x_{12}^2)}{\pi} \frac{1}{x^2_{12} } \; ,
\end{equation}
and for $x_{01} \ll x_{02}\sim x_{12}$ we
can take $\frac{\alpha_s(x_{02}^2)}{\alpha_s(x_{12}^2)} \sim 1$ and obtain
$$
{\cal K}_{rcLL}^{\rm Bal} \simeq \frac{\alpha_s(x_{01}^2) N_c}{\pi}\frac{x_{01}^2}{x^2_{12} x_{02}^2} \; .
$$

Thus for the  reference we have also used  an alternative prescription which is the minimal dipole scenario defined as 
\be
{\cal K}_{rcLL}^{\rm min} \simeq\alpha_s(\min(x_{01}^2,x_{12}^2,x_{02}^2))\frac{N_c}{\pi}\frac{x_{01}^2}{x^2_{12} x_{02}^2} \;.
\label{eq:dipmin}
\ee

  As we shall see later, even though formally kernel (\ref{eq:KernLOBal}) reduces to (\ref{eq:dipmin}), at least  in the cases considered above, there are  notable numerical differences between the two prescriptions (\ref{eq:KernLOBal},\ref{eq:dipmin}). In particular, we found that   the evolution with  kernel (\ref{eq:KernLOBal}) is significantly slower than with (\ref{eq:dipmin}) which is consistent with previous numerical results \cite{Albacete:2007yr} obtained without impact parameter.

In the case of the kernel with  running  coupling (\ref{eq:KernLOBal})  the implementation of the cutoff in the form analogous to
${\cal K}_{LL}^{(1)}$  is not entirely trivial due to the rather complicated form of (\ref{eq:KernLOBal}). We have thus used the simplest 
scenario, imposing the  cuts on the daughter dipoles as in (\ref{eq:KernLOTheta}). To be precise, for the case of the running coupling with scenario \cite{Balitsky:2006wa} we used

\beeq
{\cal K}_{rcLL,m}^{\rm Bal} = \frac{ \alpha_s(x_{01}^2) N_c}{\pi}\left[\frac{1}{x_{02}^2}\left(\frac{\alpha_s(x_{02}^2)}{\alpha_s(x_{12}^2)} - 1\right) + \frac{1}{x_{12}^2}\left(\frac{\alpha_s(x_{12}^2)}{\alpha_s(x_{02}^2)} - 1\right) + \frac{x_{01}^2}{x^2_{12} x_{02}^2}\right]\Theta(\frac{1}{m^2} - x_{02}^2)\Theta(\frac{1}{m^2} - x_{12}^2) \; .
\label{eq:KernLOBalTheta}
\eeeq

For comparison,
 we have also used the minimal dipole prescription for the running coupling (\ref{eq:dipmin}) where all of the scenarios  with masses (\ref{eq:KernLOBessMass},\ref{eq:KernLOTheta}) were extended in this case. This is possible as the minimal dipole prescription  gives the multiplicative factor in the LL kernel, see (\ref{eq:dipmin}).

   In this paper we use the expression of the QCD running coupling with a mass parameter $\mu$ to regulate the strong coupling at large  dipoles, which is of the following form\footnote{Note that, this is the expression for the running coupling in coordinate space. In the literature one finds various forms of the running of the coupling in coordinate space which include different normalizations for the argument, i.e. $\ln \frac{C}{\Lambda^2 x^2}$ with differing values of the constant $C$. We are using here the convention from Ref.~\cite{Balitsky:2008zza} where $C=1$.}

\be
\alpha_s(x^2) = \frac{1}{b \, \ln\left[\Lambda_{\rm QCD}^{-2}\left(\frac{1}{x^2} + \mu^2\right)\right]}  \; ,
\label{eq:coupling}
\ee

Here $b = \frac{33 - 2n_f}{12\pi}$, $n_f$ is the number of active flavors. The $\mu$ parameter effectively freezes the coupling at  large dipole sizes at $\alpha_{s,{\rm fr }} = \frac{1}{b \ln\left[\Lambda^{-2}\mu^2\right]}$. In our simulations we used  $\mu=0.52 \; {\rm GeV}$ as an infra-red regulator for the strong coupling, and $\Lambda_{\rm QCD}=0.25 \; {\rm GeV}$.

\begin{figure}
\centering
\subfigure[\hspace*{0.1cm}Kernel given by Eq.~(\ref{eq:KernLOBessMass}) with running coupling $\alpha_s(\min(x_{01}^2,x_{02}^2,x_{12}^2))$.]{\label{fig:mar05-2}\includegraphics[angle=270,width=0.49\textwidth]{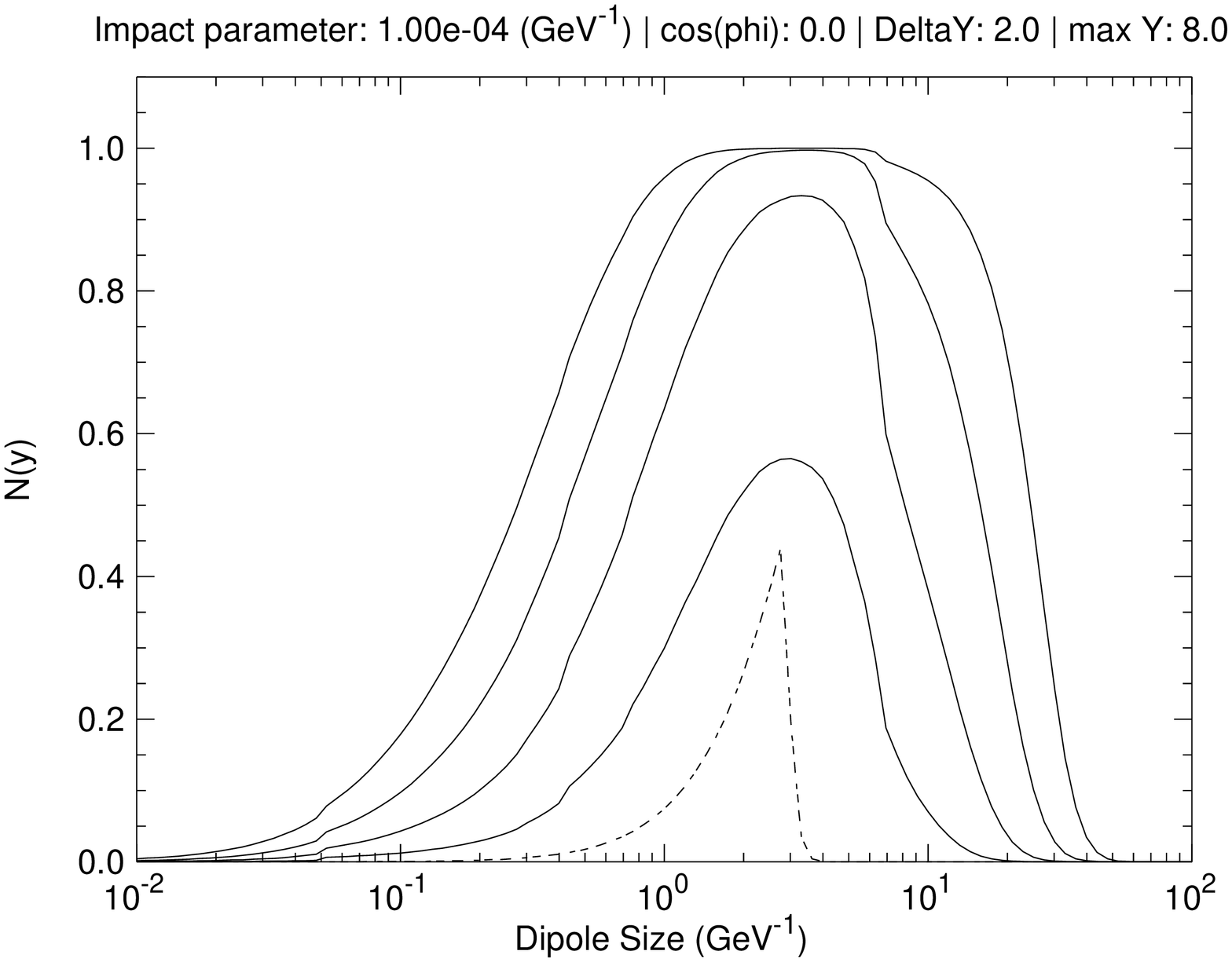}}
\subfigure[\hspace*{0.1cm}Kernel given by Eq.~(\ref{eq:KernLOTheta}) with running coupling $\alpha_s(\min(x_{01}^2,x_{02}^2,x_{12}^2))$.]{\label{fig:mar05-0}\includegraphics[angle=270,width=0.49\textwidth]{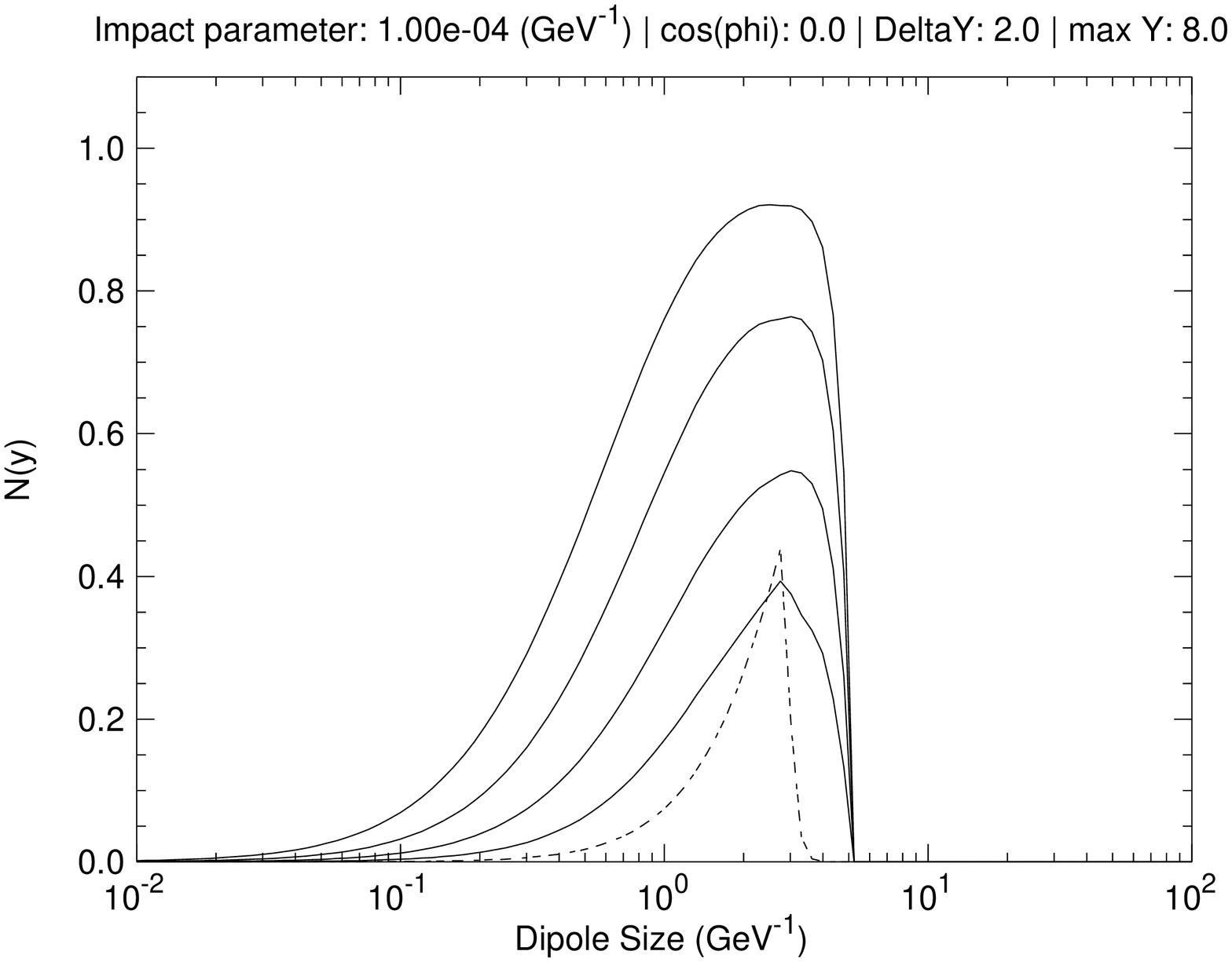}}
\caption{Dipole amplitude as a function of the dipole size from the evolution to the BK equation with mass $m=0.35 \; {\rm GeV}$.   The dotted-dashed line is the initial condition (\ref{eq:glaubermueller}) at $Y=0$ and each solid line represents a progression in two units of rapidity until $Y=8$.
The solutions are evaluated at fixed impact parameter $b=0.0001 \; {\rm GeV}^{-1}$ and for
the orientation of the dipole defined as ${\bf r} \perp {\bf b}$. }
\label{fig:mar05}
\end{figure}


\section{Results}
\label{sec:results}
\subsection{Properties of the  dipole amplitude and comparison with the Glauber-Mueller model}

As  we saw explicitly in Sec.~\ref{sec:Evol}, the solutions to the BK equation with impact parameter dependence  possess very interesting and novel  properties as compared to the case without the impact parameter. In this section we investigate in detail the  features of the solutions when the mass $m$ is included in the kernel. We study the BK solutions using two different prescriptions for the infrared regulators which were introduced in the previous section,  see  Eqs.~(\ref{eq:KernLOBessMass}) and (\ref{eq:KernLOTheta}).  We also investigate in detail the differences between the two running coupling scenarios, i.e. (\ref{eq:dipmin}) 
and  (\ref{eq:KernLOBalTheta}). It turns out that the differences between the simulations have  rather significant impact on the phenomenology.

First, we used  kernels (\ref{eq:KernLOBessMass},\ref{eq:KernLOTheta}),  where  the running coupling has been implemented using the minimal dipole size as the scale (see Eq.~\ref{eq:dipmin}). This was done in order to  consistently trace the differences between the two implementations of the massive cutoff.

In Fig.~\ref{fig:mar05-2} we present the simulations using kernel  (\ref{eq:KernLOBessMass}). 
We observe that despite the fact that the effective gluon mass $m$ is incorporated into the branching kernel, the scattering amplitude is non-vanishing at arbitrary large dipole sizes. For any rapidity there is still a significant diffusion into the large dipole size region. This feature is simple to understand by inspecting the form of (\ref{eq:KernLOBessMass}). One observes that, 
 for any value of $x_{01}$ there are configurations where one daughter dipole
  is very large and above the cutoff but the second daughter dipole
  can be still below the cutoff $\frac{1}{m}$.  These configurations lead to the  non-vanishing contributions of the kernel even at large values of $x_{01}$.  Because of this effect the evolution with modified kernel (\ref{eq:KernLOBessMass}) still proceeds into the large dipole regime as indicated by the results in Fig.~\ref{fig:mar05-2}.

In  Fig.~\ref{fig:mar05-0}  we show the solution using the  kernel with the theta functions  imposed as in (\ref{eq:KernLOTheta}).  
We see from this figure that the prescription (\ref{eq:KernLOTheta}) leads to the solution which completely vanishes for large dipoles in stark contrast with the simulation shown in Fig.~\ref{fig:mar05-2}.
The interesting feature about this scenario  is that even though the initial condition (shown by dotted-dashed line) was cut at $x_{01} = \frac{1}{m}$, the evolution does not respect this cut and moves it  to a larger value equal to $x_{01} = \frac{2}{m}$.  This happens because the kernel (\ref{eq:KernLOTheta}) has cuts only on daughter dipoles and not on the parent dipole. It means that 
the amplitude for sizes of dipoles which are larger than $\frac{1}{m}$ is non-vanishing.
  Until a point at which the parent dipole is twice the cutoff size there exist configurations where neither emitted daughter dipoles are above the cutoff size.  These symmetric states exist until $x_{01} = \frac{2}{m}$ at which point all of the configurations are cut, since at least one daughter dipole is larger than the cutoff.

From this part of the analysis we conclude that 
 when the massive cutoff is imposed in the form of  (\ref{eq:KernLOBessMass}), the evolution still proceeds into the region of large dipole sizes, which is dominated by the large value of the   strong coupling constant. This has rather important consequences and we found that this scenario is   actually disfavoured by the data.

\begin{figure}
\includegraphics[angle=270,width=0.6\textwidth]{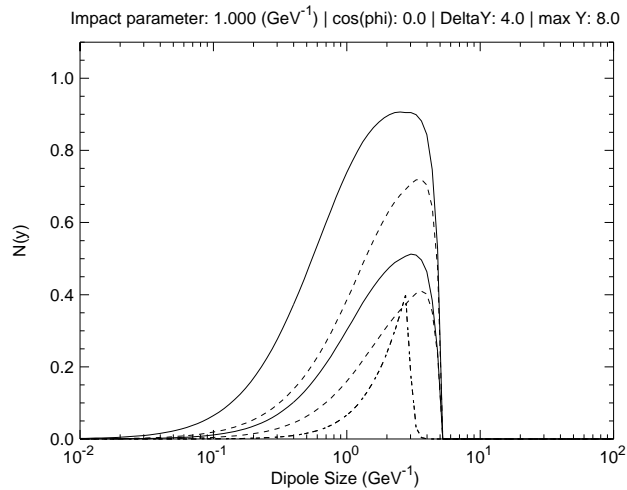}
\caption{Dipole amplitude as a function of the dipole size for the fixed value of impact parameter. Solid line corresponds to the simulation with the running coupling using  the minimum dipole prescription Eq.~(\ref{eq:dipmin}).  Dashed line corresponds to the running coupling using  Eq.~(\ref{eq:KernLOBalTheta}). In both cases  the mass  $m=0.35 \, {\rm GeV}$ is included in the kernel, in the form of two theta functions, like in  Eq.~(\ref{eq:KernLOTheta}). Two sets of curves correspond to rapidities $Y=4$ and $Y=8$.
The dashed - dotted curve is the initial condition at $Y=0$ for both calculations.}
\label{fig:MinBalComp}
\end{figure}

As a next step, we have performed the comparison of the two solutions using the  minimal dipole prescription  (\ref{eq:dipmin})  and the Balitsky prescription  (\ref{eq:KernLOBalTheta}) for the running coupling. The results of this comparison  are shown in Fig.~\ref{fig:MinBalComp}. In both of these cases we have implemented the mass as in (\ref{eq:KernLOTheta}), i.e. setting the kernel to zero whenever any of the daughter dipoles are larger than the cutoff.  It is evident that the kernel with the running coupling given by (\ref{eq:KernLOBalTheta})
leads to a much slower evolution than the kernel   with the running coupling as in (\ref{eq:dipmin}). For example at rapidity $Y=4$, the evolution front in the minimal dipole scenario is almost at the same position  (for small dipoles) as for the scenario (\ref{eq:KernLOBalTheta}) at rapidity $Y=8$. As we will see later, this will translate  into large differences for the observable structure functions depending on the prescription used.  It has a large effect, in particular on the $x$ slope of $F_2$. This feature has been found also in earlier calculation which did not include the  impact parameter dependence and  masses, see Ref.~\cite{Albacete:2007yr}.

\begin{figure}
\centering
\subfigure[\hspace*{0.1cm} { Initial  condition regulated at $\frac{1}{m}$.}]{\label{fig:BalKow1}\includegraphics[angle=270,width=0.44\textwidth]{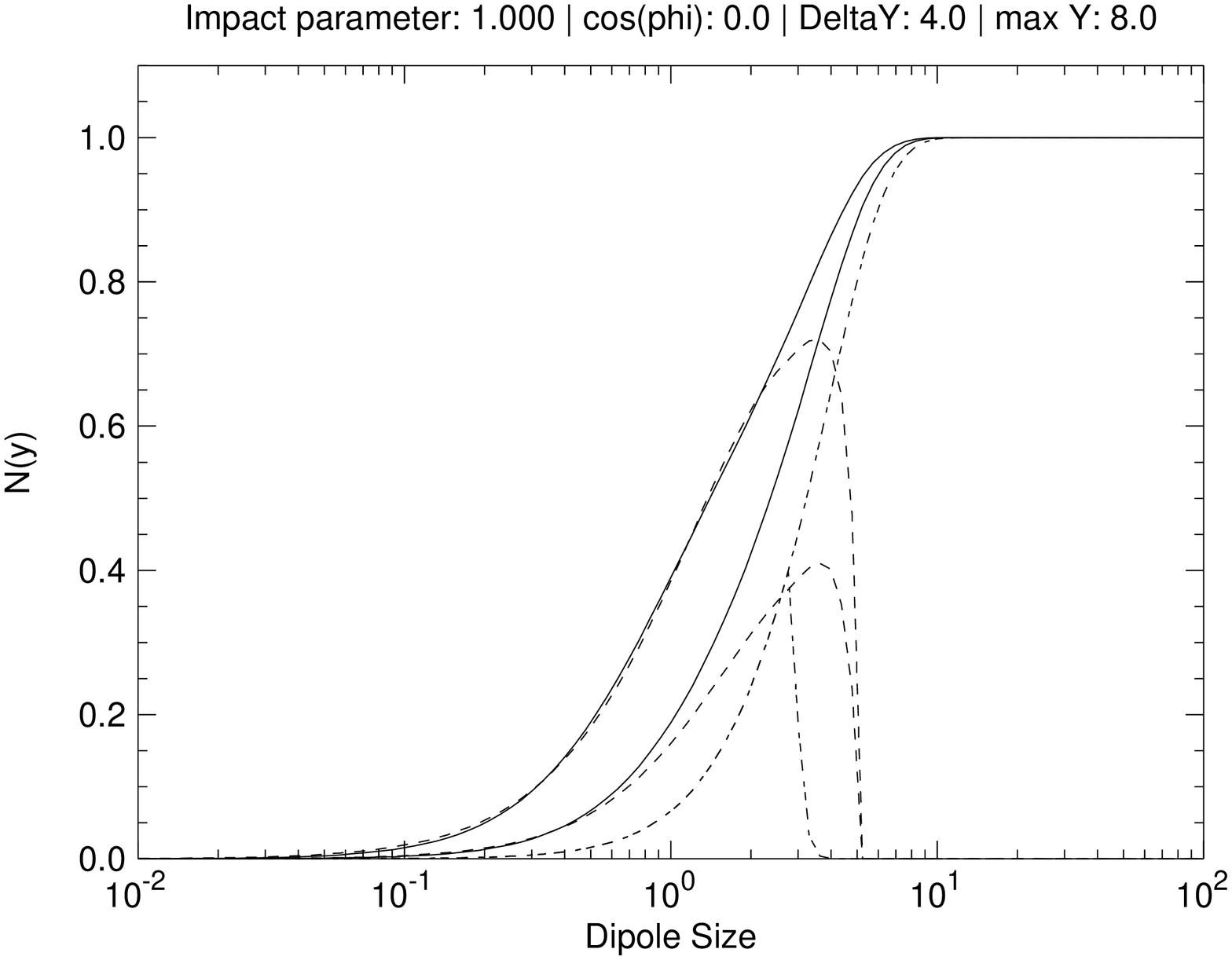}}
\subfigure[\hspace*{0.1cm} { Initial condition regulated at $\frac{2}{m}$.}]{\label{fig:BalKow2}\includegraphics[angle=270,width=0.49\textwidth]{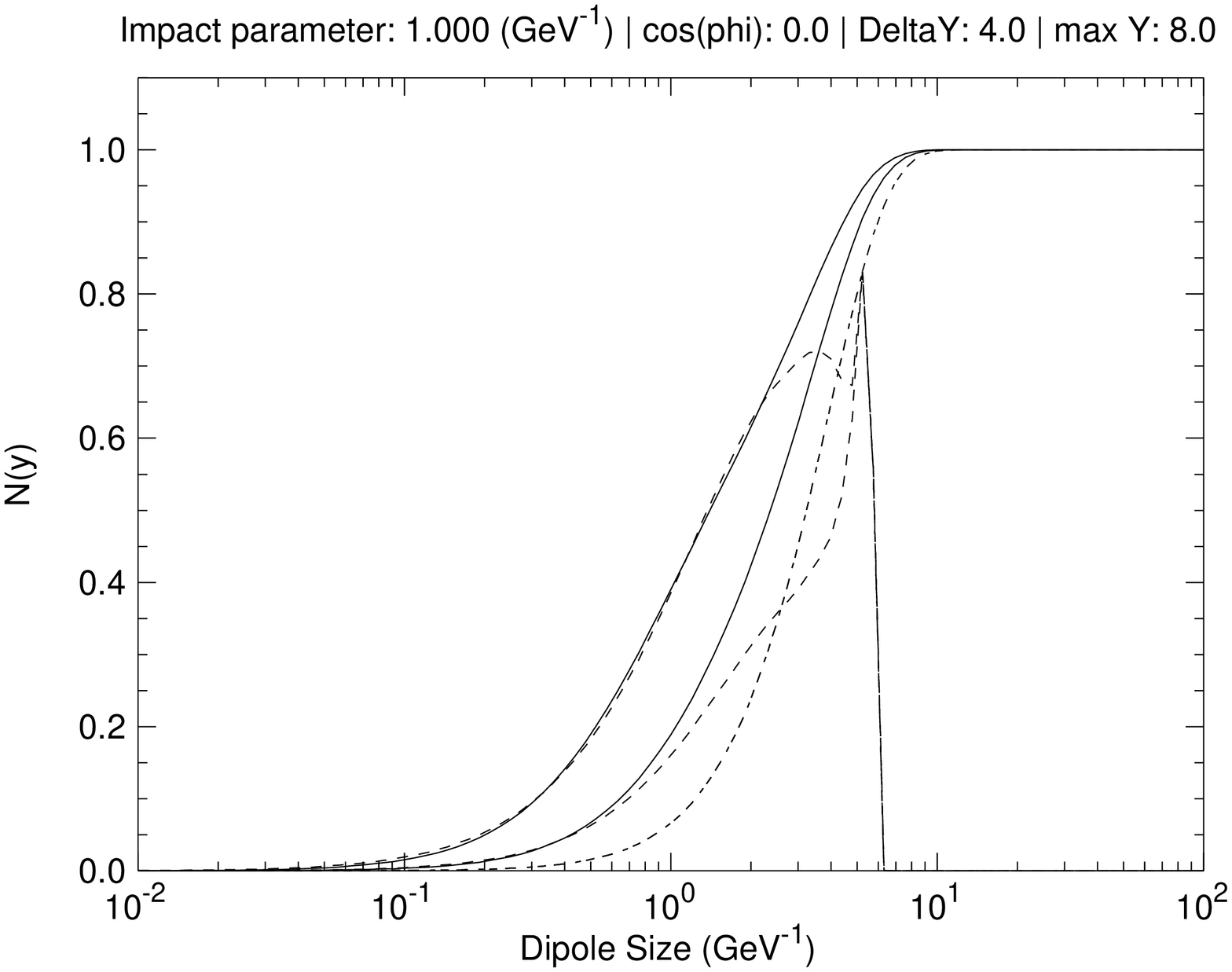}}
\caption{ Dipole scattering amplitude as a function of the dipole size for the fixed value of impact parameter $b$. Solid line corresponds to the model (\ref{eq:glaubermueller}) with parameters from \cite{Kowalski:2006hc}.
  The dashed line is the solution to the BK equation with the kernel  (\ref{eq:KernLOBalTheta}).  The dashed-dotted line denotes the model (\ref{eq:glaubermueller}) at $x_0=0.01$. The cutoff at $r_{\rm max}=\frac{1}{m}$ (plot (a)) and  $r_{\rm max}=\frac{2}{m}$ (plot (b)) is also indicated. Impact parameter is fixed to $b= 1 \; {\rm GeV}^{-1}$.}
\label{fig:BalKow}
\end{figure}

We have also compared 
  the solutions to the BK equation  with the Glauber-Mueller model, Eq.~\ref{eq:glaubermueller}. In the latter model, the parameters  were   obtained from a fit to the HERA data ~\cite{Kowalski:2006hc}. 
The initial condition for the solution to BK was also taken to be of the form of Eq.~(\ref{eq:glaubermueller})
at $Y=0$ which we choose to correspond to $x_0=0.01$ in this calculation. For consistency the initial condition for the BK equation is set to zero for dipoles which exceed the cutoff. 
The comparison between the solution to BK and Glauber-Mueller model is presented in Fig.~\ref{fig:BalKow1}.
We see that the solution to the BK equation agrees quite well with the parametrization (\ref{eq:glaubermueller}) for small values of dipole sizes.  On the other hand  there are sizeable differences in the larger dipole regime, where one sees that  (\ref{eq:glaubermueller}) extends indefinitely whereas the BK solution is cut off by the massive regulator. This has a non-negligible impact on the phenomenology of $F_2$ at low $Q^2$ 
as will be illustrated in the next subsection.

The difference in the solutions was examined, when the initial condition is not cut at $\frac{1}{m}$ but at $\frac{2}{m}$  and still the kernel (\ref{eq:KernLOBalTheta}) is used with the cutoff $\frac{1}{m}$ in the evolution. The result is shown in Fig.~\ref{fig:BalKow2}. In this case it is evident that the cutoff is not moved due to the reasons described earlier in this section.  However,  there is a peculiar structure of the solution, where  a second peak of the amplitude emerges for dipoles somewhat smaller than the cutoff.  The solution in Fig.~\ref{fig:BalKow2} is nevertheless very close to the solution shown in Fig.~\ref{fig:BalKow1} for small values of dipole sizes, i.e. smaller than $\frac{1}{m}$.
We also note that for the values of the cutoff $m$  used here, which are motivated by the profile in impact parameter, both the initial condition  and the BK solutions do not completely saturate, although  the amplitude is close to $1$ for large rapidities. 

It should be also noted that  the solutions to the BK equation with the kernel (\ref{eq:KernLOBal}) possesses an  interesting dependence on the regularization procedure of the running coupling. This is related to the fact that this kernel is a complicated, non-linear function of $\alpha_s$. In particular, the factors of the coupling  in the denominators in expression(\ref{eq:KernLOBal})  lead to a non-trivial dependence on the regularization parameter of the strong coupling.
  It was found that by  increasing the value of  $\mu$, see Eq.(\ref{eq:coupling}), and thus decreasing the maximal value at which the coupling freezes,  there was a region  of dipole sizes where the solution obtained from evolution with the kernel  (\ref{eq:KernLOBal}) was actually increased, contrary to what could be naively expected.   We stress that this behavior was observed for some range of dipole sizes only. With the minimal dipole size prescription the amplitude was of course always decreasing with increasing value of $\mu$ as expected.

In general, it was found that the solution with the running coupling in the form (\ref{eq:KernLOBal}) possessed a larger sensitivity to the way the coupling is regulated than the evolution with the minimal dipole prescription (\ref{eq:dipmin}).
This sensitivity persists even
 at small dipole sizes which are far away from the scale $\frac{1}{\mu}$. 
 It suggests that the terms with inverse coupling in  kernel (\ref{eq:KernLOBal}) increase the sensitivity to scales which are different than the scale set by the parent dipole $x_{01}$.    It is also important to note that this behavior was found  both in the evolution with and without impact parameter dependence.  The solution using the minimum dipole size prescription does not exhibit such a large sensitivity.  It would be interesting to investigate these features further in order to determine whether this is a  physical behavior  or it is an artefact of the truncation of the resummation of perturbative series which lead to this result \cite{Balitsky:2006wa}.

\begin{figure}
\centering
\subfigure[\hspace*{0.2cm} Dipole size $r=1.0 \; {\rm GeV}^{-1}$. Logarithmic horizontal axis.]{\label{fig:bdeplog}\includegraphics[angle=270,width=0.49\textwidth]{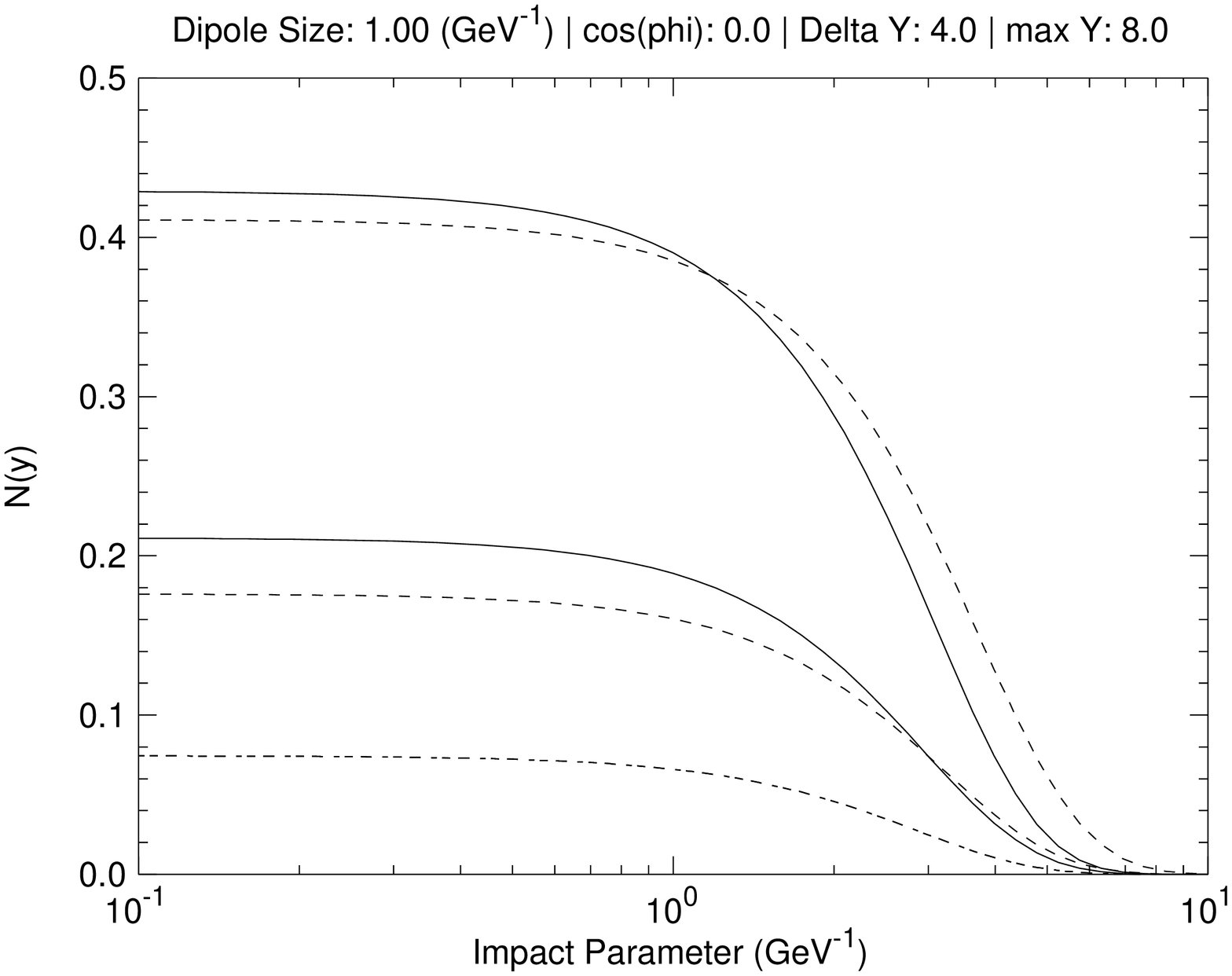}}
\subfigure[\hspace*{0.2cm} Dipole size $r=1.0 \; {\rm GeV}^{-1}$. Linear horizontal axis.]{\label{fig:bdeplin}\includegraphics[angle=270,width=0.49\textwidth]{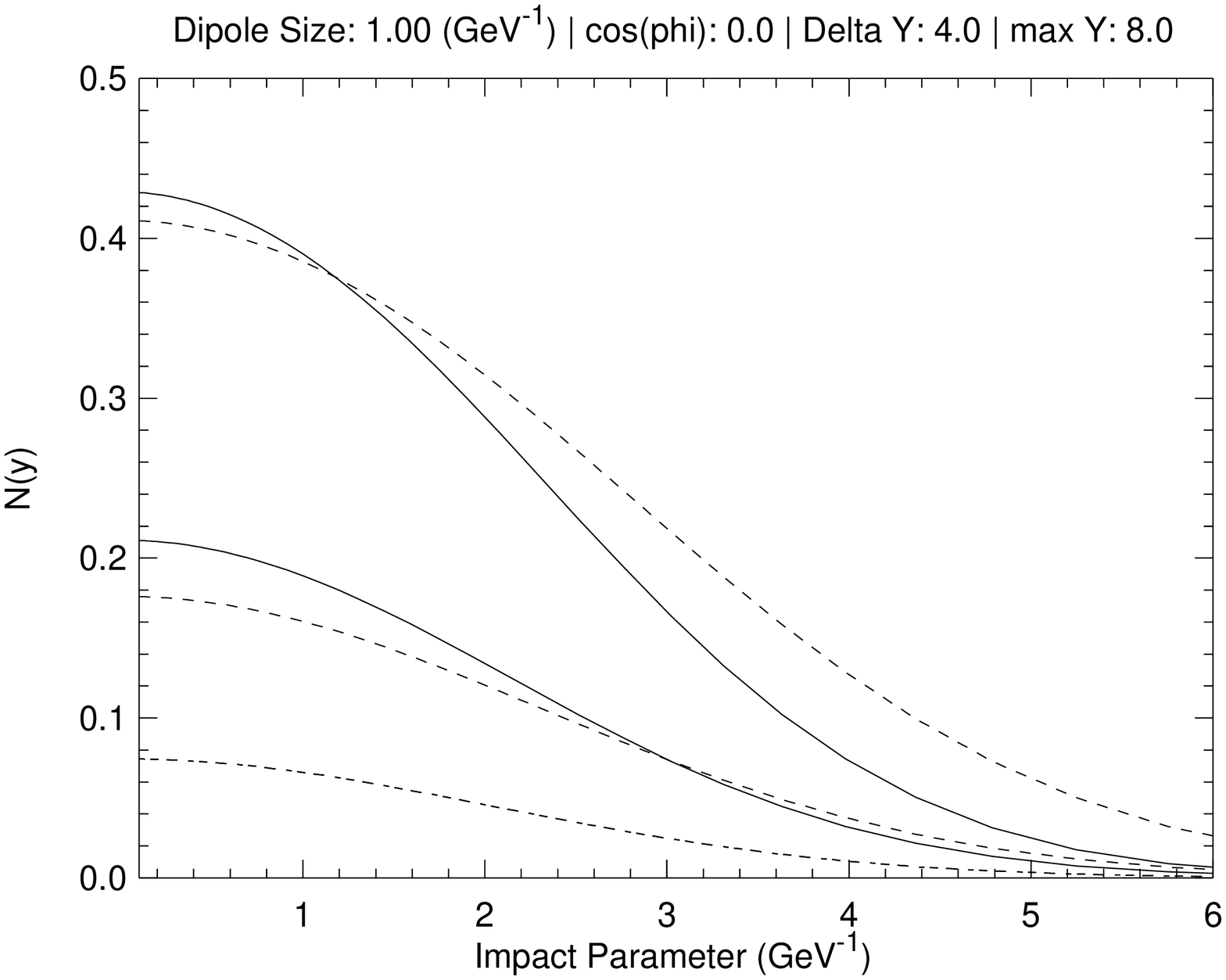}}
\caption{Dipole scattering amplitude as a function of the  impact parameter for fixed dipole size and dipole orientation $\theta=\pi/2$.  The solid lines represent the model (\ref{eq:glaubermueller}) used in \cite{Kowalski:2006hc}. The dashed lines correspond to the solution of the BK equation with the kernel (\ref{eq:KernLOBalTheta}), $m = 0.35\;{\rm GeV}$. The dashed - dotted line represents the initial conditions at $Y=0 \, (x_0=0.01)$ also taken from model in \cite{Kowalski:2006hc}.}
\label{fig:bdep}
\end{figure}

Finally, the dependence of the amplitude on impact parameter for fixed value of the dipole size was analyzed.
The diffusion property of the solution in impact parameter space  is illustrated in Fig.~\ref{fig:bdep} for the running coupling case in scenario (\ref{eq:KernLOBalTheta}).  
Plots in Fig.\ref{fig:bdeplog} and Fig.~\ref{fig:bdeplin} differ only by the choice of horizontal scale. 
 The solution to the BK equation  is compared with the profile in impact parameter using the model (\ref{eq:glaubermueller}) with the parameters from \cite{Kowalski:2006hc}. We observe that for the small values of $b$ in general the BK solution is fairly close numerically to model (\ref{eq:glaubermueller}). There is however  a significant  difference in the shape of the amplitudes between the BK calculation and the Glauber-Mueller model (\ref{eq:glaubermueller}). This is especially manifest at large values of impact parameter where the BK solution has a more extended tail in $b$. 
In the BK solution there is a clear increase of the width of the distribution in impact parameter with increasing rapidity.

\begin{figure}
\centering
\includegraphics[angle=270,width=0.6\textwidth]{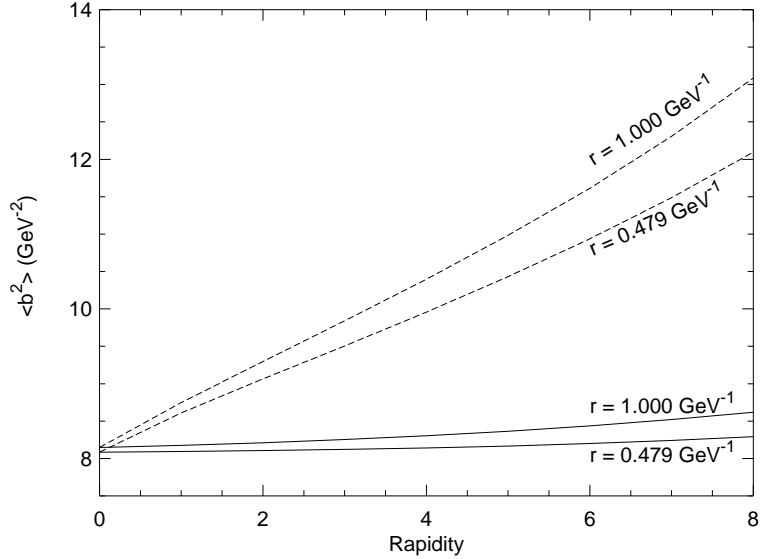}
\caption{The value of the average squared width $\langle b^2 \rangle$, defined in Eq.~(\ref{eq:bvariance}), as a function of rapidity for fixed value of the dipole size $r$. The solid line is the model (\ref{eq:glaubermueller}) with parameters taken from \cite{Kowalski:2006hc} and the dashed line is obtained from solution to the BK  equation with the kernel (\ref{eq:KernLOBalTheta}).}
\label{fig:baverage}
\end{figure}

The diffusion property in impact parameter is best illustrated in Fig.~\ref{fig:baverage}
where we show the average width squared, defined as
\be
\langle b^2 \rangle \; = \; \frac{\int d^2 \bb \, b^2 \, N(\rb,\bb;Y)}{\int d^2 \bb \, N(\rb,\bb;Y)} \; ,
\label{eq:bvariance}
\ee
as a function of rapidity for fixed value of the dipole size $r$.
We compared the value of $\langle b^2 \rangle$ extracted from the solution to the BK equation with the value obtained from model (\ref{eq:glaubermueller}). The model (\ref{eq:glaubermueller})
gives almost constant width, independent of rapidity, which is to be expected. On the contrary,  in  the case of the BK equation the width clearly increases with rapidity. For the rapidities considered here, we observe that it is almost a linear growth, with slightly faster increase at the highest values of rapidity $\sim 6-8$ along with mild dependence of the slope on the value of the dipole size. 

\subsection{Description of the $F_2$ and $F_L$ structure function data}

In our calculation of the $F_2$ and $F_L$ structure functions we used the solution to the BK equation with impact parameter dependence and a gluon mass $m$ as implemented in scenario  (\ref{eq:KernLOBalTheta}). The initial condition for the calculation   was set as  in Sec.~\ref{sec:initial}. In addition, we imposed a   cutoff on the dipole sizes to be equal to   $x_{01} = \frac{2}{m}$ in the initial condition. As discussed earlier, this is necessary in order to be consistent with the cutoff placed in the evolution kernel.
 The data on the structure function correspond to the combined H1 and ZEUS data \cite{:2009wt}, and only the points below $x=0.01$ are used here.

In this calculation there are the following parameters: $C$, $\eta_0$, $B_G$ in the  initial condition (\ref{eq:glaubermueller}), mass $m$ in the kernel, strong coupling regulator $\mu$ and $\Lambda_{QCD}$. In principle all of these parameters can be varied to obtain the fit to the data. In practice however, due to very long time needed to find solution for a given set of parameters (of the order of a day on 32 cores), variation of all parameters is prohibitive. We have instead chosen to fix all of the parameters with the exception of $\mu$ and $m$ which we varied in a very limited range.

Since we use an initial condition which is cut at large dipole sizes, the data at values of $x$ around $0.01$ are underestimated by our model. This is because the initial model (\ref{eq:glaubermueller}) from \cite{Kowalski:2006hc} was fitted to the data without any cuts. 
This indicates a rather large sensitivity of the $F_2$ obtained from (\ref{eq:F2}) to the region of large dipole sizes even for moderate values of $Q^2$. This effect is  well known and corresponds to the presence of the aligned jet configurations in the transverse structure function  \cite{Golec-Biernat:1998js,Ewerz:2011ph}.  This can be seen by inspecting Eqs.~(\ref{eq:F2},\ref{eq:PhotonT}) as this expression receives large contributions from the endpoints $z\sim 0,1$. As a result, at a given value of $Q^2$ the dipoles which contribute to $F_2$ form a rather wide distribution in dipole size. We have verified that for the model given by  (\ref{eq:glaubermueller})  with a cut of the order of $\frac{1}{m}$ and $\frac{2}{m}$, with $m=0.35 \; {\rm GeV}$ the contribution to $F_2$ at $Q^2= 4.5 \; {\rm GeV}^2$ from dipoles larger than the cut is about $30\%$ and $10\%$ respectively.
  Hence, the $F_2$ structure function contains significant  non-perturbative contributions even at the moderate values of $Q^2$.  In order to compensate for this non-perturbative off-set one could of course move the cutoff $m$ towards smaller masses. However, since the cut on the dipole sizes is strongly correlated with the profile in impact parameter it would result in the much larger width of the impact parameter profile, which would be inconsistent with the data on diffractive $J/\Psi$ production. Therefore, we choose to work with the value of the cut which is more consistent with the number obtained from the $J/\Psi$ diffractive slope.
As a result, in order to compensate for the offset, a separate non-perturbative contribution is added which is important at low values of $Q^2$, of the order of $< 15 \; {\rm GeV}^2$.

 We stress that this property stems from the fact that in the BK evolution equation the impact parameter and dipole size are strongly correlated. This has to be contrasted with the Glauber-Mueller like parametrization (\ref{eq:glaubermueller}) where the dipole sizes and impact parameter are  decoupled. 
The non-perturbative part originating from large dipole sizes was parametrized in the following form

\be
F_{2}^{\rm soft} = \frac{Q^2}{2 \pi \alpha_{em}} \sigma_0 \int_{\frac{1}{m}} r \,dr\int_0^1 dz \left(|\Psi_L|^2 + |\Psi_T|^2\right) \; .
\label{eq:F2soft}
\ee
The total structure function is then taken to be of the form
\be
F_{2}^{\rm tot} = F_2^{\rm BK} + F_2^{\rm soft} \; ,
\label{eq:F2tots}
\ee
where $F_2^{\rm BK}$ denotes the contribution obtained by using the solution to the BK equation (which is the perturbative part).
In the formula (\ref{eq:F2soft}) we assumed that the dominant part of the integral is where the dipole - proton amplitude is almost flat and therefore replaced the dipole cross section with the constant $\sigma_0$. Note that the integral over dipole size in (\ref{eq:F2soft}) is now cut from below by $1/m$, and the integral extends into the large dipole regime.
$\sigma_0$ is a constant  that is used to fit the data at $x=0.01$ and lowest bin in $Q^2$. 
 The $F_2^{\rm soft}$ contribution is slowly varying with $Q^2$ and it accounts well for the non-perturbative dipoles at low $Q^2$. This procedure of adding separate contributions from small (perturbative) and large (non-perturbative) dipoles is similar to the  one employed in Refs.~\cite{Kwiecinski:1987tb,Martin:1998dr}.

One  could argue that the non-perturbative contribution could be accounted for by including the contribution from the vector meson dominance model  \cite{Brodsky:1969iz,Gribov:1968gs,Ritson:1970yu,Sakurai:1972wk}. The vector meson dominance (VMD) contribution  can be written as 

\be
F_{2}^{(\rm VMD)} = \frac{Q^2}{4\pi}\Sigma_{v=\rho,\omega,\phi} \left(\frac{m_v^4 \sigma_v(W^2)}{\gamma^2_v (Q^2 + m_v^2)^2}+\frac{Q^2 m_v^2 \sigma_v(W^2)}{\gamma_v^2 (Q^2 + m_v^2)^2}\xi_0\left(\frac{m_v^2}{Q^2 + m_v^2}\right) ^2\right) \; ,
\label{eq:F2vmd}
\ee
see for example \cite{Kwiecinski:1987tb}.
Here $m_v$ denotes the vector meson mass and $\sigma_v$ is the vector meson-proton cross section which is a function of the energy. These cross sections can be taken 
 to be 
\begin{eqnarray}
\sigma_\rho &=& \sigma_\omega = \frac{1}{2} \left(\sigma(\pi^+p)+\sigma(\pi^-p)\right) \; ,
\\\sigma_\phi &=& \sigma(K^+p) + \sigma(K^-p) -\frac{1}{2} \left(\sigma(\pi^+p)+\sigma(\pi^-p)\right) \; ,
\end{eqnarray}

where the parameterization for the energy dependence of $\sigma(\pi^{\pm}p),\sigma(K^{\pm}p)$ was taken using the soft pomeron model from \cite{Donnachie:1992ny}.  The $\gamma_v$ terms relate to the leptonic width of the vector meson \cite{Kwiecinski:1987tb} in question and are defined by $\gamma_v^2 = \frac{\pi \alpha_{em}^2 m_v}{3 \Gamma_{v\rightarrow e^+e^-}}$ where the leptonic decay width is taken from \cite{Nakamura:2010zzi} and $\xi_0$ is a constant taken to be $0.7$ \cite{Kwiecinski:1987tb}.

\begin{figure}
\centering
\includegraphics[angle=270,width=0.9\textwidth]{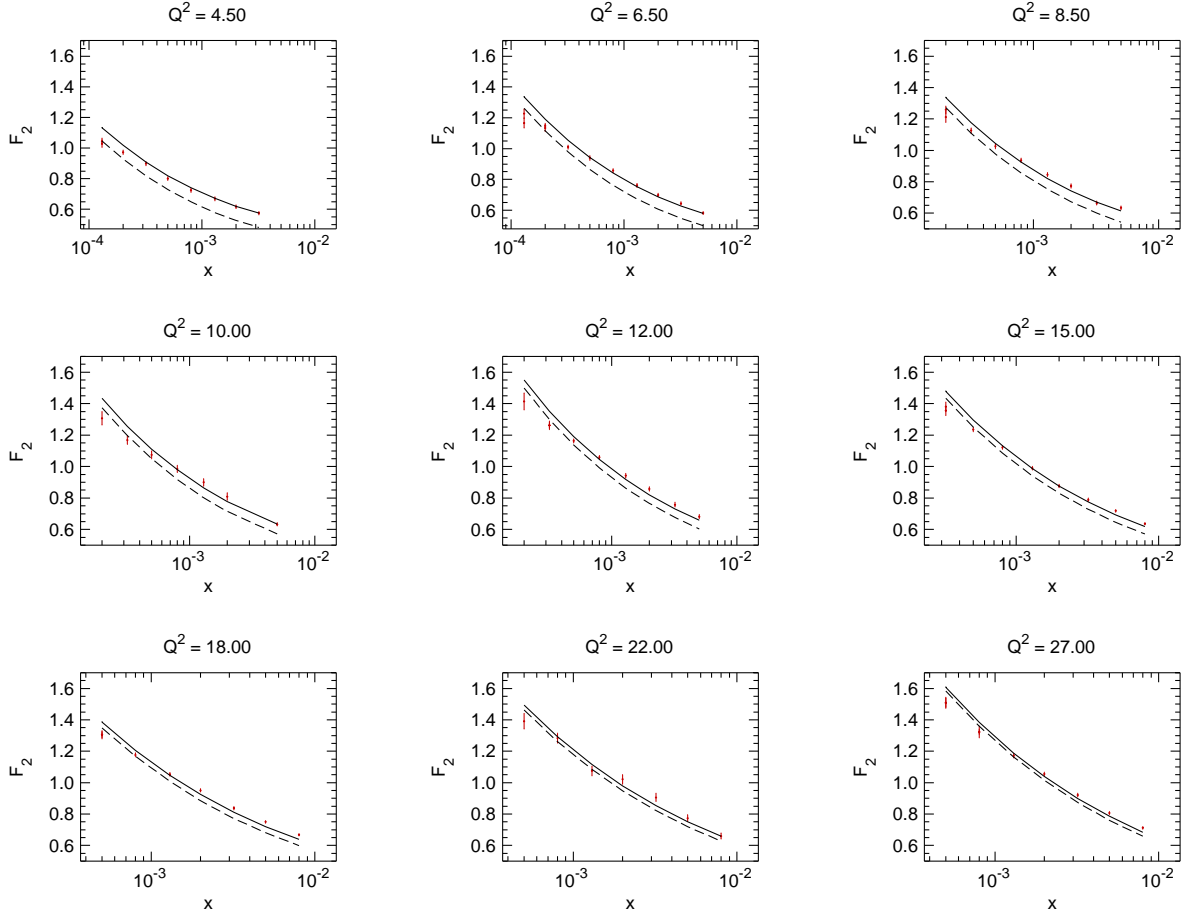}
\caption{$F_2$ proton structure function versus $x$ in bins of $Q^2$. The dipole amplitude is obtained from the BK evolution with LL kernel with running coupling  in scenario (\ref{eq:KernLOBalTheta}).  $\mu = 0.52 \; {\rm GeV}$, $m = 0.35 \; {\rm GeV}$, $\sigma_0 = 75.98 \; {\rm GeV}^{-2}$ and the initial condition is cut at $x_{01} = \frac{2}{m}$.  The dashed line corresponds to the calculation with the VMD term in $F_2$ and the solid line corresponds to the calculation (\ref{eq:F2soft},\ref{eq:F2tots}) .  The data points are taken from the combined H1 and ZEUS data sets \cite{:2009wt}.}
\label{fig:mar21-0F2NT1}
\end{figure}

\begin{figure}
\centering
\includegraphics[angle=270,width=0.9\textwidth]{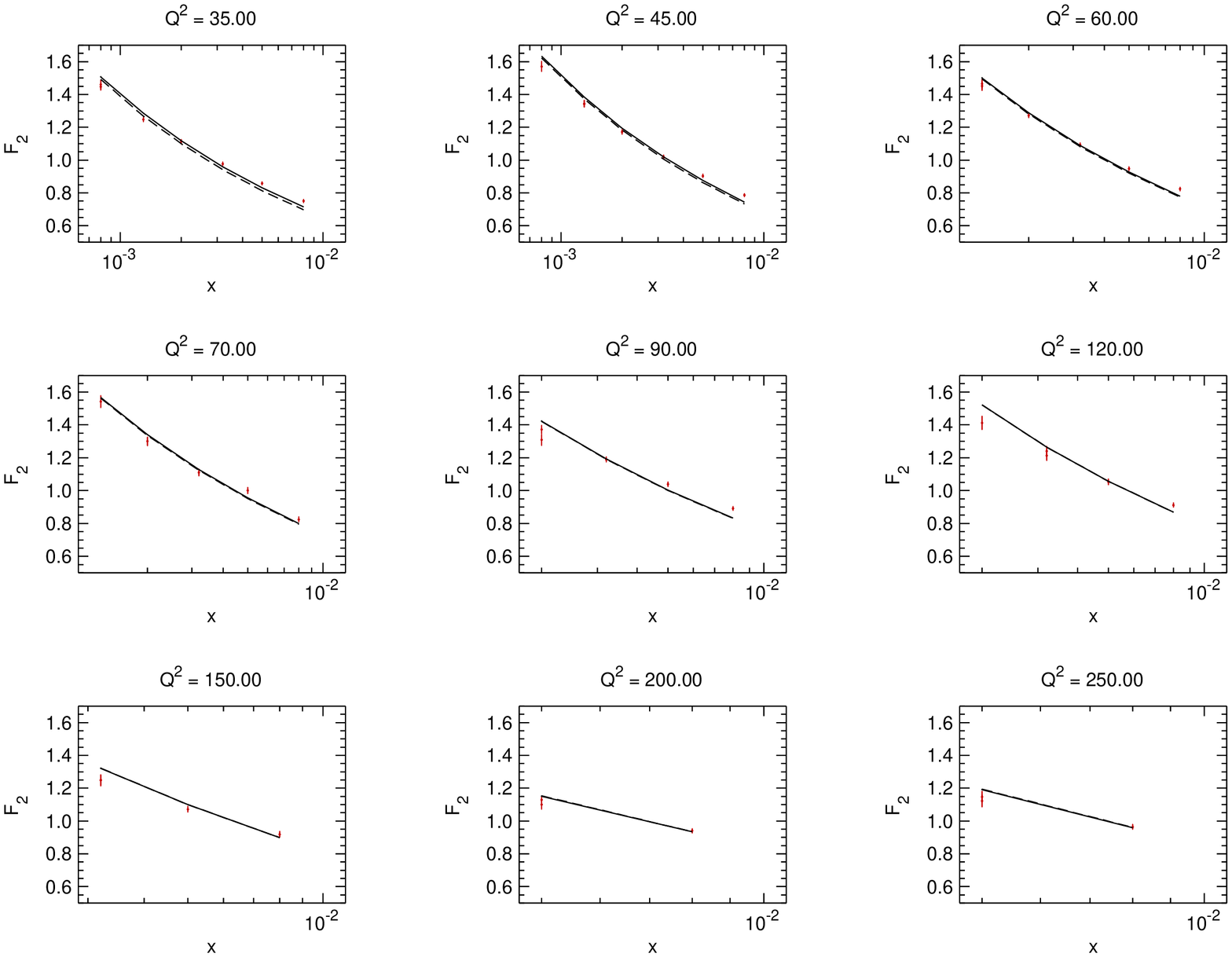}
\caption{$F_2$ proton structure function versus $x$ in bins of $Q^2$. The dipole amplitude is obtained from the BK evolution with LL kernel with running coupling  in scenario (\ref{eq:KernLOBalTheta}).  $\mu = 0.52 \; {\rm GeV}$, $m = 0.35 \; {\rm GeV}$, $\sigma_0 =  75.98 \; {\rm GeV}^{-2}$ and the initial condition is cut at $x_{01} = \frac{2}{m}$.  The dashed line corresponds to the calculation with the VMD term in $F_2$ and the solid line corresponds to the calculation (\ref{eq:F2soft},\ref{eq:F2tots}) .  The data points are taken from the combined H1 and ZEUS data sets \cite{:2009wt}.}
\label{fig:mar21-0F2NT2}
\end{figure}

The computations for $F_2$ using the non-perturbative contribution of either (\ref{eq:F2soft}) or (\ref{eq:F2vmd}) are presented in Figs.~\ref{fig:mar21-0F2NT1},\ref{fig:mar21-0F2NT2}.
As we see the curves with the VMD term undershoot the data whereas the curves with the soft term are systematically closer. The VMD contribution is almost negligible with the exception of the lowest $Q^2$ bin.  We see that the two calculations, with soft term and VMD are close for values of $Q^2>15 \, {\rm GeV}^2$ which means in this region the data are only described by the perturbative component. We also conclude that the VMD model is not sufficient as the only soft contribution for the description of the data in the present setup.  VMD only contributes to the very low values of $Q^2$ (less than $4 \; {\rm GeV}^2$) and an additional  soft component is needed  which has a flatter $Q^2$ dependence. This result is consistent with the results of the  previous analyses \cite{Martin:1998dr}.

We observe that the  slope of the calculations  is too steep in $x$ in all bins of $Q^2$ which implies that the LL evolution with the running coupling leads to a  faster slope in $\ln 1/x$ than the data indicate.  This can be remedied by lowering the scale $\Lambda_{QCD}$ from the value which we used, i.e. $ 0.25\, {\rm GeV}$, and taking  it as a fitting parameter.
This was effectively done in the fits presented in \cite{Albacete:2009fh,Albacete:2010sy}. We estimated that it  would require $\Lambda_{QCD}$ to be well below the pion mass, of the order of tens of $\rm MeV$ or so to fit the data.  
The fact that the LL evolution with running coupling in our scenario has a steeper slope than the data is  not unexpected as one needs to take into account the next-to-leading corrections to BK equation. These have been computed in \cite{Balitsky:2008zza}, but a detailed analysis of the BK equation which includes them still needs to be performed. Preliminary analysis in the momentum space  was recently performed  \cite{Avsar:2011ds} using the method of the saturation boundary\cite{Mueller:2002zm}. The results from this analysis indicate that next-to-leading corrections to the BK equation, which are not due to the running coupling, are indeed substantial and can lead to the instabilities of the evolution despite the presence of saturation \cite{Beuf:2010aw}. This strongly suggests  that a resummation of subleading corrections in $\ln 1/x$ is needed in addition to the saturation corrections \cite{Motyka:2009gi,Beuf:2011}.

\begin{figure}
\centering
\includegraphics[angle=270,width=0.9\textwidth]{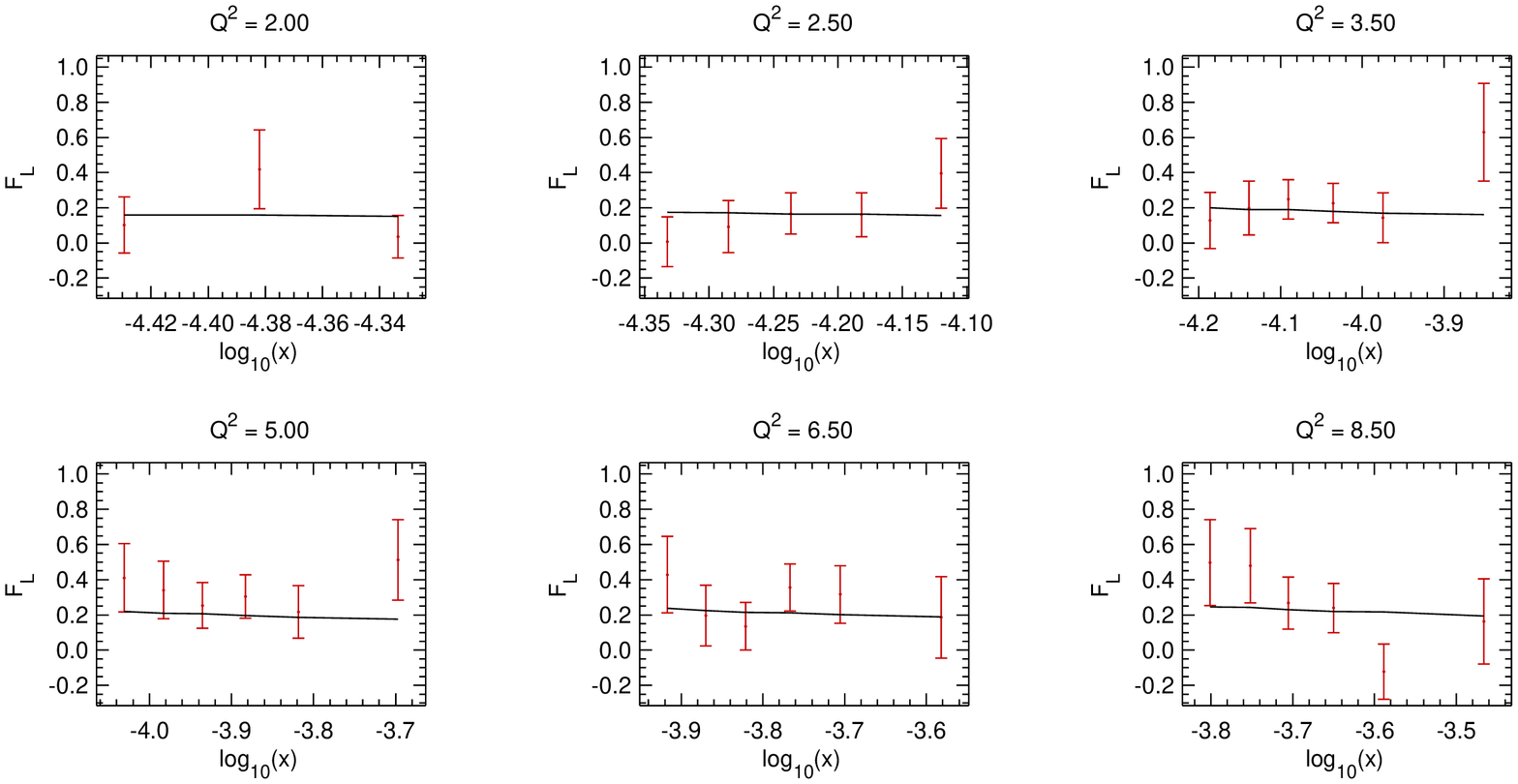}
\includegraphics[angle=270,width=0.9\textwidth]{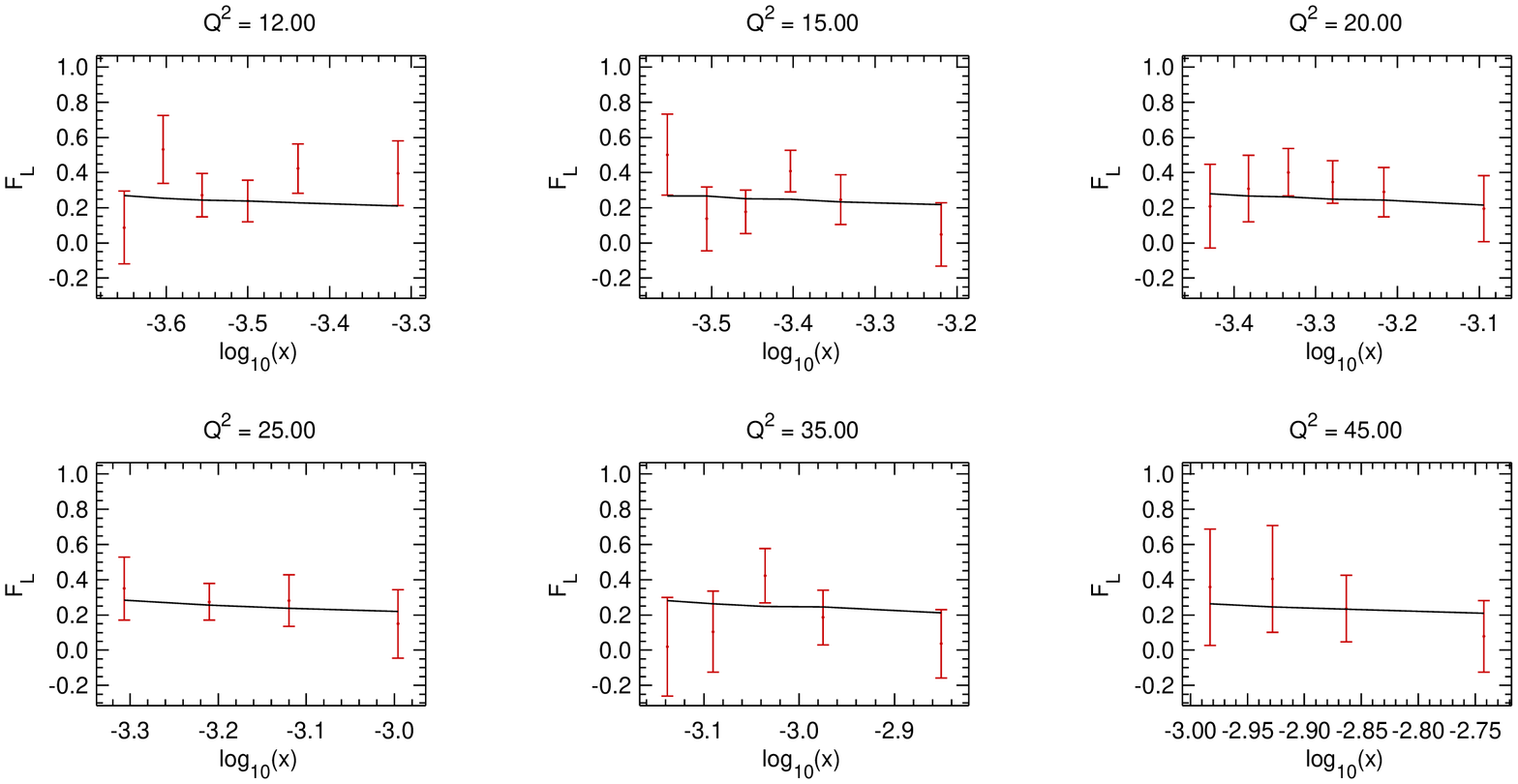}
\caption{$F_L$ proton structure function versus $x$ in bins of $Q^2$. The dipole amplitude is obtained from the BK evolution with LL kernel with running coupling  in scenario (\ref{eq:KernLOBalTheta}).  The parameters are: $\mu = 0.52 \; {\rm GeV}$, $m = 0.35 \; {\rm GeV}$. The initial condition is cut at $x_{01} = \frac{2}{m}$.  The non-perturbative contributions are not included as they are much smaller for the longitudinal structure function. Note the horizontal axis range changes from bin to bin and in generally contains very small range in $x$.}
\label{fig:mar07-0FL}
\end{figure}

Finally, we also compared the calculation to the experimental data on the structure function $F_L$ \cite{Collaboration:2010ry}, this is illustrated in Fig.~\ref{fig:mar07-0FL}.
We see that the calculation  is consistent with the experimental data in all bins of $Q^2$. However, the data on $F_L$ have very large errors.  Note that,  in the figures presented, the range in $x$ in each of the bins is very small, and therefore the $F_L$ structure function is very flat in each of these bins.

\section{Conclusions}
\label{sec:conclusion}

In this paper we have analyzed the nonlinear BK equation with impact parameter dependence and running coupling in the presence of a mass scale, which regulates dipole splitting in the infrared regime. This effective gluon mass is responsible for the non-perturbative effect of confinement. Using the resulting solution for the dipole scattering amplitude we have  performed a comparison with experimental DIS data on the structure functions $F_2$ and $F_L$.  Let us summarize the main points of this investigation :

\begin{enumerate}
\item The  details of the evolution in rapidity strongly depend on the way the large dipoles are regularized. In particular, the speed of the evolution with rapidity is  affected by the choice of regularization. Two different scenarios have been tested: 
modified Bessel functions (\ref{eq:KernLOBessMass}), and 
a more stringent cutoff with theta functions (\ref{eq:KernLOTheta}). 
The first scenario possesses 
a  physical  motivation and can be derived from the  computation of the branching kernel for dipoles in the presence of the effective gluon mass in the propagators.

\item The scenario  (\ref{eq:KernLOBessMass})
does not entirely tame the evolution into the large dipole size region and in the presence of the running coupling it results in a rather fast evolution in rapidity.
The resulting amplitude is also much larger than in scenario (\ref{eq:KernLOTheta}), not only in the large dipole regime but also in the small dipole region as well. We found that the scenario  with the  cutoff on all the large dipoles in the form  (\ref{eq:KernLOTheta})   results in solutions which are preferred  by the experimental data.

\item The running coupling prescription (\ref{eq:KernLOBal}) gives, in general, much slower evolution than the  prescription (\ref{eq:dipmin}) with the minimal dipole as the scale of the running coupling. This happens despite the fact that the two kernels are  formally equivalent in the limits when the dipole sizes are strongly ordered. We also found that the evolution with kernel (\ref{eq:KernLOBal}) is more dependent on the details of the regularization of the running coupling. Most likely this is caused by the highly nonlinear form of (\ref{eq:KernLOBal}), i.e. by  the fact that it contains the inverse powers of the strong coupling. We found that the scenario which gives results closest to the data is the kernel with Balitsky prescription for the running coupling and the massive regulator taken in  the form (\ref{eq:KernLOBalTheta}).
 
\item Comparison with the data on structure function $F_2$ shows that the slope in $x$ of the calculation is  too steep for the data in the case of the LL evolution with running coupling. This could be cured by  changing the value of  $\Lambda_{QCD}$, but it would require unrealistically small values of this scale in order to match the data. We stress that the full fit which would include the variation of all parameters,  in the case of the BK equation with impact parameter, would be extremely demanding as far as computing resources are concerned (using similar techniques as presented here). The fact that the scale in the running coupling has to be adjusted quite a bit to fit the data  is consistent with previous calculations (which were done without impact parameter dependence) and calls for the inclusion of the remaining corrections beyond the leading - logarithmic order.

\item An important feature of the solution to the BK equation is the fact that the impact parameter and dipole size are strongly correlated. This property  stems from the basic form of the dipole evolution kernel. This has to be contrasted with the models previously used in the literature, for example of the form   (\ref{eq:glaubermueller}).  This correlation introduces novel effects and leads to more constraints on the calculations. For example,  incorporating  the effective gluon mass in the kernel introduces a scale in impact parameter.  Strong variation of this scale is not possible if one requires consistency with the observed slope in diffractive data.  This scale then results in the truncation of the large dipole size region, which in turn causes the offset in the $F_2$ calculation, particularly at  small values of $Q^2$. Therefore one needs to include an additional (soft) component to $F_2$ which is entirely non-perturbative. The BK solution also exhibits the diffusion in impact parameter, a feature that is completely absent in the models of the form (\ref{eq:glaubermueller}). We found that in the region of large impact parameters and high rapidities the differences between the BK solution and the model (\ref{eq:glaubermueller}) were  substantial.
 
\end{enumerate}

\section*{Acknowledgments}
We would like to thank Emil Avsar, Dionysis Triantafyllopoulos and Yuri Kovchegov for many useful discussions. We also thank Henri Kowalski for discussions as well as his assistance by allowing us usage of parts of his fortran code for the evaluation of the initial conditions. This work was supported  by the  MNiSW grant No. N202 249235  and the DOE OJI grant No. DE - SC0002145.  A.M.S. is supported by the Sloan Foundation.

\bibliographystyle{h-physrev4}
\bibliography{mybib}

\begin{thebibliography}{10}

\bibitem{Deshpande:2005wd}
A.~Deshpande, R.~Milner, R.~Venugopalan and W.~Vogelsang,
\newblock Ann.Rev.Nucl.Part.Sci. {\bf 55}, 165 (2005), [hep-ph/0506148].

\bibitem{Boer:2011fh}
D.~Boer {\em et~al.},
\newblock 1108.1713.

\bibitem{Dainton:2006wd}
J.~Dainton, M.~Klein, P.~Newman, E.~Perez and F.~Willeke,
\newblock JINST {\bf 1}, P10001 (2006), [hep-ex/0603016].

\bibitem{Klein:2009zz}
M.~Klein and P.~Newman,
\newblock CERN Cour. {\bf 49N3}, 22 (2009).

\bibitem{Gribov:1984tu}
L.~V. Gribov, E.~M. Levin and M.~G. Ryskin,
\newblock Phys. Rept. {\bf 100}, 1 (1983).

\bibitem{Mueller:1985wy}
A.~H. Mueller and J.-w. Qiu,
\newblock Nucl.Phys. {\bf B268}, 427 (1986).

\bibitem{Kovchegov:1999yj}
Y.~V. Kovchegov,
\newblock Phys. Rev. {\bf D60}, 034008 (1999), [hep-ph/9901281].

\bibitem{Kovchegov:1999ua}
Y.~V. Kovchegov,
\newblock Phys. Rev. {\bf D61}, 074018 (2000), [hep-ph/9905214].

\bibitem{Balitsky:1995ub}
I.~Balitsky,
\newblock Nucl. Phys. {\bf B463}, 99 (1996), [hep-ph/9509348].

\bibitem{Balitsky:1998ya}
I.~Balitsky,
\newblock Phys. Rev. {\bf D60}, 014020 (1999), [hep-ph/9812311].

\bibitem{Balitsky:2001re}
I.~Balitsky,
\newblock Phys. Lett. {\bf B518}, 235 (2001), [hep-ph/0105334].

\bibitem{Balitsky:1998kc}
I.~Balitsky,
\newblock Phys. Rev. Lett. {\bf 81}, 2024 (1998), [hep-ph/9807434].

\bibitem{McLerran:1993ka}
L.~D. McLerran and R.~Venugopalan,
\newblock Phys. Rev. {\bf D49}, 3352 (1994), [hep-ph/9311205].

\bibitem{McLerran:1993ni}
L.~D. McLerran and R.~Venugopalan,
\newblock Phys. Rev. {\bf D49}, 2233 (1994), [hep-ph/9309289].

\bibitem{JalilianMarian:1997gr}
J.~Jalilian-Marian, A.~Kovner, A.~Leonidov and H.~Weigert,
\newblock Phys. Rev. {\bf D59}, 014014 (1999), [hep-ph/9706377].

\bibitem{JalilianMarian:1997dw}
J.~Jalilian-Marian, A.~Kovner and H.~Weigert,
\newblock Phys. Rev. {\bf D59}, 014015 (1999), [hep-ph/9709432].

\bibitem{Weigert:2000gi}
H.~Weigert,
\newblock Nucl. Phys. {\bf A703}, 823 (2002), [hep-ph/0004044].

\bibitem{Iancu:2000hn}
E.~Iancu, A.~Leonidov and L.~D. McLerran,
\newblock Nucl. Phys. {\bf A692}, 583 (2001), [hep-ph/0011241].

\bibitem{Iancu:2001ad}
E.~Iancu, A.~Leonidov and L.~D. McLerran,
\newblock Phys. Lett. {\bf B510}, 133 (2001), [hep-ph/0102009].

\bibitem{Ferreiro:2001qy}
E.~Ferreiro, E.~Iancu, A.~Leonidov and L.~McLerran,
\newblock Nucl. Phys. {\bf A703}, 489 (2002), [hep-ph/0109115].

\bibitem{Mueller:2001uk}
A.~H. Mueller,
\newblock Phys. Lett. {\bf B523}, 243 (2001), [hep-ph/0110169].

\bibitem{Dumitru:2011vk}
A.~Dumitru, J.~Jalilian-Marian, T.~Lappi, B.~Schenke and R.~Venugopalan,
\newblock 1108.4764.

\bibitem{Fadin:1975cb}
V.~S. Fadin, E.~A. Kuraev and L.~N. Lipatov,
\newblock Phys. Lett. {\bf B60}, 50 (1975).

\bibitem{Balitsky:1978ic}
I.~I. Balitsky and L.~N. Lipatov,
\newblock Sov. J. Nucl. Phys. {\bf 28}, 822 (1978).

\bibitem{Lipatov:1985uk}
L.~N. Lipatov,
\newblock Sov. Phys. JETP {\bf 63}, 904 (1986).

\bibitem{Levin:1999mw}
E.~Levin and K.~Tuchin,
\newblock Nucl.Phys. {\bf B573}, 833 (2000), [hep-ph/9908317].

\bibitem{Levin:2000mv}
E.~Levin and K.~Tuchin,
\newblock Nucl.Phys. {\bf A691}, 779 (2001), [hep-ph/0012167].

\bibitem{Mueller:2002zm}
A.~H. Mueller and D.~N. Triantafyllopoulos,
\newblock Nucl. Phys. {\bf B640}, 331 (2002), [hep-ph/0205167].

\bibitem{Munier:2003sj}
S.~Munier and R.~B. Peschanski,
\newblock Phys. Rev. {\bf D69}, 034008 (2004), [hep-ph/0310357].

\bibitem{Braun:2001kh}
M.~A. Braun,
\newblock hep-ph/0101070.

\bibitem{Lublinsky:2001yi}
M.~Lublinsky, E.~Gotsman, E.~Levin and U.~Maor,
\newblock Nucl. Phys. {\bf A696}, 851 (2001), [hep-ph/0102321].

\bibitem{Armesto:2001fa}
N.~Armesto and M.~A. Braun,
\newblock Eur. Phys. J. {\bf C20}, 517 (2001), [hep-ph/0104038].

\bibitem{Lublinsky:2001bc}
M.~Lublinsky,
\newblock Eur. Phys. J. {\bf C21}, 513 (2001), [hep-ph/0106112].

\bibitem{GolecBiernat:2001if}
K.~J. Golec-Biernat, L.~Motyka and A.~M. Stasto,
\newblock Phys. Rev. {\bf D65}, 074037 (2002), [hep-ph/0110325].

\bibitem{Rummukainen:2003ns}
K.~Rummukainen and H.~Weigert,
\newblock Nucl.Phys. {\bf A739}, 183 (2004), [hep-ph/0309306].

\bibitem{Albacete:2007yr}
J.~L. Albacete and Y.~V. Kovchegov,
\newblock Phys. Rev. {\bf D75}, 125021 (2007), [0704.0612].

\bibitem{Albacete:2009fh}
J.~L. Albacete, N.~Armesto, J.~G. Milhano and C.~A. Salgado,
\newblock Phys.Rev. {\bf D80}, 034031 (2009), [0902.1112].

\bibitem{Albacete:2010sy}
J.~Albacete, N.~Armesto, J.~Milhano, P.~Arias and C.~Salgado,
\newblock 1012.4408.

\bibitem{Albacete:2007sm}
J.~L. Albacete,
\newblock Phys.Rev.Lett. {\bf 99}, 262301 (2007), [0707.2545].

\bibitem{Albacete:2010bs}
J.~L. Albacete and C.~Marquet,
\newblock Phys.Lett. {\bf B687}, 174 (2010), [1001.1378].

\bibitem{Albacete:2010fs}
J.~L. Albacete,
\newblock J.Phys.Conf.Ser. {\bf 270}, 012052 (2011), [1010.6027].

\bibitem{ALbacete:2010ad}
J.~L. ALbacete and A.~Dumitru,
\newblock 1011.5161.

\bibitem{Munier:2001nr}
S.~Munier, A.~Stasto and A.~H. Mueller,
\newblock Nucl.Phys. {\bf B603}, 427 (2001), [hep-ph/0102291].

\bibitem{Kowalski:2003hm}
H.~Kowalski and D.~Teaney,
\newblock Phys.Rev. {\bf D68}, 114005 (2003), [hep-ph/0304189].

\bibitem{Kowalski:2006hc}
H.~Kowalski, L.~Motyka and G.~Watt,
\newblock Phys.Rev. {\bf D74}, 074016 (2006), [hep-ph/0606272].

\bibitem{Salam:1995zd}
G.~P. Salam,
\newblock Nucl. Phys. {\bf B449}, 589 (1995), [hep-ph/9504284].

\bibitem{Salam:1995uy}
G.~P. Salam,
\newblock Nucl. Phys. {\bf B461}, 512 (1996), [hep-ph/9509353].

\bibitem{Avsar:2005iz}
E.~Avsar, G.~Gustafson and L.~Lonnblad,
\newblock JHEP {\bf 07}, 062 (2005), [hep-ph/0503181].

\bibitem{Avsar:2006gw}
E.~Avsar,
\newblock Acta Phys. Polon. {\bf B37}, 3561 (2006), [hep-ph/0610045].

\bibitem{Avsar:2006jy}
E.~Avsar, G.~Gustafson and L.~Lonnblad,
\newblock JHEP {\bf 01}, 012 (2007), [hep-ph/0610157].

\bibitem{Avsar:2007xh}
E.~Avsar,
\newblock JHEP {\bf 11}, 027 (2007), [0709.1371].

\bibitem{GolecBiernat:2003ym}
K.~J. Golec-Biernat and A.~M. Stasto,
\newblock Nucl. Phys. {\bf B668}, 345 (2003), [hep-ph/0306279].

\bibitem{Berger:2010sh}
J.~Berger and A.~Stasto,
\newblock Phys. Rev. {\bf D83}, 034015 (2011), [1010.0671].

\bibitem{Gotsman:2004ra}
E.~Gotsman, M.~Kozlov, E.~Levin, U.~Maor and E.~Naftali,
\newblock Nucl. Phys. {\bf A742}, 55 (2004), [hep-ph/0401021].

\bibitem{Gubser:2011qva}
S.~S. Gubser,
\newblock 1102.4040.

\bibitem{Collaboration:2010ry}
F.~Aaron {\em et~al.},
\newblock Eur.Phys.J. {\bf C71}, 1579 (2011), [1012.4355].

\bibitem{:2009wt}
H1 and ZEUS Collaboration, F.~Aaron {\em et~al.},
\newblock JHEP {\bf 1001}, 109 (2010), [0911.0884].

\bibitem{Nikolaev:1990ja}
N.~N. Nikolaev and B.~Zakharov,
\newblock Z.Phys. {\bf C49}, 607 (1991).

\bibitem{Nikolaev:1991et}
N.~Nikolaev and B.~G. Zakharov,
\newblock Z.Phys. {\bf C53}, 331 (1992).

\bibitem{Kovner:2001bh}
A.~Kovner and U.~A. Wiedemann,
\newblock Phys. Rev. {\bf D66}, 051502 (2002), [hep-ph/0112140].

\bibitem{Kovner:2002yt}
A.~Kovner and U.~A. Wiedemann,
\newblock Phys. Lett. {\bf B551}, 311 (2003), [hep-ph/0207335].

\bibitem{Dudal:2010tf}
D.~Dudal, O.~Oliveira and N.~Vandersickel,
\newblock Phys.Rev. {\bf D81}, 074505 (2010), [1002.2374].

\bibitem{Oliveira:2010xc}
O.~Oliveira and P.~Bicudo,
\newblock J.Phys.G {\bf G38}, 045003 (2011), [1002.4151].

\bibitem{Flensburg:2008ag}
C.~Flensburg, G.~Gustafson and L.~Lonnblad,
\newblock Eur.Phys.J. {\bf C60}, 233 (2009), [0807.0325].

\bibitem{Aktas:2005xu}
H1, A.~Aktas {\em et~al.},
\newblock Eur. Phys. J. {\bf C46}, 585 (2006), [hep-ex/0510016].

\bibitem{BerStaDiff}
J.~Berger and A.~Stasto,
\newblock work in progress.

\bibitem{Nikolaev:1993th}
N.~N. Nikolaev and B.~Zakharov,
\newblock Z.Phys. {\bf C64}, 631 (1994), [hep-ph/9306230].

\bibitem{Nikolaev:1993ke}
N.~N. Nikolaev, B.~Zakharov and V.~Zoller,
\newblock JETP Lett. {\bf 59}, 6 (1994), [hep-ph/9312268].

\bibitem{Mueller:1993rr}
A.~H. Mueller,
\newblock Nucl. Phys. {\bf B415}, 373 (1994).

\bibitem{Balitsky:2006wa}
I.~Balitsky,
\newblock Phys. Rev. {\bf D75}, 014001 (2007), [hep-ph/0609105].

\bibitem{Kovchegov:2006vj}
Y.~V. Kovchegov and H.~Weigert,
\newblock Nucl. Phys. {\bf A784}, 188 (2007), [hep-ph/0609090].

\bibitem{Balitsky:2008zza}
I.~Balitsky and G.~A. Chirilli,
\newblock Phys. Rev. {\bf D77}, 014019 (2008), [0710.4330].

\bibitem{Golec-Biernat:1998js}
K.~Golec-Biernat and M.~Wusthoff,
\newblock Phys. Rev. {\bf D59}, 014017 (1999), [hep-ph/9807513].

\bibitem{Ewerz:2011ph}
C.~Ewerz, A.~von Manteuffel and O.~Nachtmann,
\newblock JHEP {\bf 1103}, 062 (2011), [1101.0288].

\bibitem{Kwiecinski:1987tb}
J.~Kwiecinski and B.~M. Badelek,
\newblock Z.Phys. {\bf C43}, 251 (1989).

\bibitem{Martin:1998dr}
A.~D. Martin, M.~Ryskin and A.~Stasto,
\newblock Eur.Phys.J. {\bf C7}, 643 (1999), [hep-ph/9806212].

\bibitem{Brodsky:1969iz}
S.~J. Brodsky and J.~Pumplin,
\newblock Phys.Rev. {\bf 182}, 1794 (1969).

\bibitem{Gribov:1968gs}
V.~Gribov,
\newblock Sov.Phys.JETP {\bf 30}, 709 (1970).

\bibitem{Ritson:1970yu}
D.~Ritson,
\newblock Phys.Rev. {\bf D3}, 1267 (1971).

\bibitem{Sakurai:1972wk}
J.~Sakurai and D.~Schildknecht,
\newblock Phys.Lett. {\bf B40}, 121 (1972).

\bibitem{Donnachie:1992ny}
A.~Donnachie and P.~Landshoff,
\newblock Phys.Lett. {\bf B296}, 227 (1992), [hep-ph/9209205].

\bibitem{Nakamura:2010zzi}
Particle Data Group, K.~Nakamura {\em et~al.},
\newblock J.Phys.G {\bf G37}, 075021 (2010).

\bibitem{Avsar:2011ds}
E.~Avsar, A.~M. Stasto, D.~N. Triantafyllopoulos and D.~Zaslavsky,
\newblock 1107.1252.

\bibitem{Beuf:2010aw}
G.~Beuf,
\newblock 1008.0498.

\bibitem{Motyka:2009gi}
L.~Motyka and A.~M. Stasto,
\newblock Phys. Rev. {\bf D79}, 085016 (2009), [0901.4949].

\bibitem{Beuf:2011}
G.~Beuf,
\newblock private communication.

\end{thebibliography}

\newpage

\end{document}